\documentclass[10pt]{article}

\usepackage{amsmath}
\usepackage{amssymb}  

\usepackage{bm,amsbsy}

\usepackage{graphicx}

\usepackage{cite}

\usepackage{color} 

\usepackage[labelfont=bf,labelsep=period,justification=raggedright]{caption}

\makeatletter
\renewcommand{\@biblabel}[1]{\quad#1.}
\makeatother

\date{}

\newcommand{\figref}[1]{Fig.~\ref{fig:#1}}

\newcommand\parderiv[2]{\frac{\partial #1}{\partial #2}}
\newcommand\parderivtwo[3]{\frac{\partial^2 #1}{\partial #2 \partial #3}}

\newcommand{\kl}[2]{D_{KL\!}\Big( #1 \,\Big|\!\Big|\, #2\Big)}
\newcommand{\vs}{{\mathbf{s}}} 
\newcommand{\vx}{{\mathbf{x}}}  

\newcommand{\vr}{{\mathbf{r}}}

\newcommand{\vk}{{\mathbf{k}}}
\newcommand{\vone}{\mathbf{1}}
\newcommand{\vlam}{{\bm{\lambda}}}
\newcommand{\valph}{{\bm{\alpha}}}
\newcommand{\vphi}{{\bm{\phi}}}

\newcommand{\Iss}{I_{ss}} 
\newcommand{\Issempir}{\widehat{I}_{ss}} 
\newcommand{\Iber}{I_{Ber}} 
\newcommand{\Iberempir}{\widehat{I}_{Ber}} 
 
\newcommand{\Icount}{I_{count}} 
\newcommand{\Icountempir}{\widehat{I}_{count}} 

\newcommand{\pstim}{p(\vs)}
\newcommand{\nsp}{n_{sp}}
\newcommand{\nphi}{{n_{\phi}}}
\newcommand{\nrpt}{{n_{rpt}}}
\newcommand{\nt}{{n_{t}}}
\newcommand{\pcondsp}{p(\vs|spike)}

\newcommand{\phat}{\widehat p}
\newcommand{\vphat}{\mathbf{\widehat p}}
\newcommand{\vqhat}{\mathbf{\widehat q}}
\newcommand{\qhat}{\widehat q}
\newcommand{\fhat}{\widehat f}  
\newcommand{\ghat}{\widehat g}
\newcommand{\spike}{{spike}}
\newcommand{\indic}[1]{\mathbf{1}_{#1}}
\newcommand{\trp}{^\top} 

\newcommand{\rmax}{r_{\text{max}}}
\newcommand{\jj}{^{(j)}}

\newcommand{\LL}{\mathcal{L}}
\newcommand{\Nrm}{\mathcal{N}}
\newcommand{\Kmid}{\widehat K_{MID}}
\newcommand{\argmax}[1]{\underset{#1}{\arg\max}\;}

\renewcommand{\eqref}[1]{eq.~\ref{eq:#1}}

\definecolor{darkred}{rgb}{0.75, 0, 0}
\definecolor{darkgreen}{rgb}{0,.7, 0}
\definecolor{darkblue}{rgb}{0, 0,0.5}
\definecolor{gray}{RGB}{0.5,0.5,0.5}

\begin{document}

\begin{flushleft} {\Large \textbf{The equivalence of information-theoretic
      and likelihood-based methods for neural dimensionality
      reduction} }
\\
Ross S. Williamson$^{1,2,\ast,\ast\ast}$, 
Maneesh Sahani$^{1}$, 
Jonathan W. Pillow$^{3,\ast\ast}$
\\
{\bf{1.}} Gatsby Computational Neuroscience Unit, University College London,
London, UK
\\
{\bf{2.}} Centre for Mathematics and Physics in the Life Sciences and
Experimental Biology, University College London, London, UK
\\
{\bf{3.}} Princeton Neuroscience Institute, Department of Psychology,
Princeton University, Princeton, New Jersey, USA
\\
$\ast$ Current Affiliation: Eaton-Peabody Laboratories, Massachusetts Eye
and Ear Infirmary, Boston, Massachusetts, USA \&
Center for Computational Neuroscience and Neural Technology, Boston
University, Boston, Massachusetts, USA
\\
\vspace{0.1in}
$\ast\ast$ Corresponding Authors: ross\_williamson@meei.harvard.edu, pillow@princeton.edu
\end{flushleft}

\section*{Abstract}
Stimulus dimensionality-reduction methods in neuroscience seek to
identify a low-dimensional space of stimulus features that affect a
neuron's probability of spiking.  One popular method, known as
maximally informative dimensions (MID), uses an information-theoretic
quantity known as ``single-spike information'' to identify this space.
Here we examine MID from a model-based perspective.  We show that MID
is a maximum-likelihood estimator for the parameters of a
linear-nonlinear-Poisson (LNP) model, and that the empirical
single-spike information corresponds to the normalized log-likelihood
under a Poisson model. This equivalence implies that MID does not
necessarily find maximally informative stimulus dimensions when
spiking is not well described as Poisson. We provide several examples
to illustrate this shortcoming, and derive a lower bound on the
information lost when spiking is Bernoulli in discrete time bins.  To
overcome this limitation, we introduce model-based dimensionality
reduction methods for neurons with non-Poisson firing statistics, and
show that they can be framed equivalently in likelihood-based or
information-theoretic terms. Finally, we show how to overcome
practical limitations on the number of stimulus dimensions that MID
can estimate by constraining the form of the non-parametric
nonlinearity in an LNP model.  We illustrate these methods with
simulations and data from primate visual cortex.

\section*{Author Summary}
A popular approach to the neural coding problem is to identify a
low-dimensional linear projection of the stimulus space that preserves
the aspects of the stimulus that affect a neuron's probability of
spiking.  Previous work has focused on both information-theoretic and
likelihood-based estimators for finding such projections.  Here, we show
that these two approaches are in fact equivalent.  We show that
maximally informative dimensions (MID), a popular
information-theoretic method for dimensionality reduction, is
identical to the maximum-likelihood estimator for a particular
linear-nonlinear encoding model with Poisson spiking.  One implication
of this equivalence is that MID may not find the
information-theoretically optimal stimulus projection when spiking is
non-Poisson, which we illustrate with a few simple examples.  Using
these insights, we propose novel dimensionality-reduction methods that
incorporate non-Poisson spiking, and suggest new parametrizations that
allow for tractable estimation of high-dimensional subspaces.

\section*{Introduction}

The neural coding problem, an important topic in systems and
computational neuroscience, concerns the probabilistic relationship
between environmental stimuli and neural spike responses.
Characterizing this relationship is difficult in general because of
the high dimensionality of natural signals.  A substantial literature
therefore has focused on dimensionality reduction methods for
identifying which stimuli affect a neuron's probability of firing.
The basic idea is that many neurons compute their responses in a low
dimensional subspace, spanned by a small number of stimulus features.
By identifying this subspace, we can more easily characterize the
nonlinear mapping from stimulus features to spike responses
\cite{deRuyter88,Aguera03a,Aguera03b,Simoncelli04,Bialek05}.

Neural dimensionality-reduction methods can be coarsely divided into
three classes: (1) moment-based estimators, such as spike-triggered
average (STA) and covariance (STC)
\cite{deRuyter88,Chichilnisky01,Bialek05,Schwartz06,Saleem08}; (2)
model-based estimators, which rely on explicit forward encoding models
\cite{Brillinger88,Kass01,Paninski04,Truccolo05,Pillow08,ParkI11,McFarland13,Cui13};
and (3) information and divergence-based estimators, which seek to
reduce dimensionality using an information-theoretic cost function
\cite{Paninski03a,Sharpee04,Pillow06,Kouh09,Fitzgerald11,Rajan13plos}.
For all such methods, the goal is to find a set of linear filters,
specified by the columns of a matrix $K$, such that the probability of
response $r$ given a stimulus $\vs$ depends only on the linear
projection of $\vs$ onto these filters, i.e.,
$p(r|\vs) \approx p(r|K\trp\vs)$.
Existing methods differ in computational complexity, modeling
assumptions, and stimulus requirements.  Typically, moment-based
estimators have low computational cost but succeed only for restricted
classes of stimulus distributions, whereas information-theoretic and
likelihood-based estimators allow for the arbitrary stimuli but have
high computational cost.  Previous work has established theoretical
connections between moment-based and likelihood-based estimators
\cite{Paninski03a,Paninski04,Pillow06,ParkI11,ParkI13gqm}, and between
some classes of likelihood-based and information-theoretic estimators
\cite{Kouh09,Fitzgerald11,ParkI11,Rajan13}.

Here we focus on maximally informative dimensions (MID), a well-known
information-theoretic estimator introduced by Sharpee, Rust \& Bialek
\cite{Sharpee04}.  We show that this estimator is formally identical
to the maximum likelihood (ML) estimator for the parameters of a
linear-nonlinear-Poisson (LNP) encoding model.  Although previous work
has demonstrated an asymptotic equivalence between these methods
\cite{Kinney07,Kouh09,Rajan13}, we show that the correspondence is
exact, regardless of time bin size or the amount of data.  This
equivalence follows from the fact that the plugin estimate for the
single-spike information \cite{Brenner00b}, the quantity that MID
optimizes, is equal to a normalized Poisson log-likelihood.

The connection between the MID estimator and the LNP model makes clear
that MID does not incorporate information carried by non-Poisson
statistics of the response.  We illustrate this shortcoming by showing
that MID can fail to find information-maximizing filters for simulated
neurons with binary or other non-Poisson spike count distributions.
To overcome this limitation, we introduce new dimensionality-reduction
estimators based on non-Poisson noise models, and show that they can
be framed equivalently in information-theoretic or likelihood-based
terms.

Finally, we show that a model-based perspective leads to strategies
for overcoming a limitation of traditional MID, that it cannot
tractably estimate more than two or three filters. The difficulty
arises from the intractability of using histograms to estimate
densities in high-dimensional subspaces.  However, the single-spike
information depends only on the ratio of densities, which is
proportional to the nonlinearity in the LNP model. We show that by
restricting the parametrization of this nonlinearity so that the
number of parameters does not grow exponentially with the number of
dimensions, we can obtain flexible yet computationally tractable
estimators for models with many filters or dimensions.

\section*{Results}

\subsection*{Background}

\subsubsection*{Linear-nonlinear-Poisson (LNP) encoding model}

\begin{figure}[!t]
  \begin{center}
    \includegraphics[width=0.75\columnwidth]{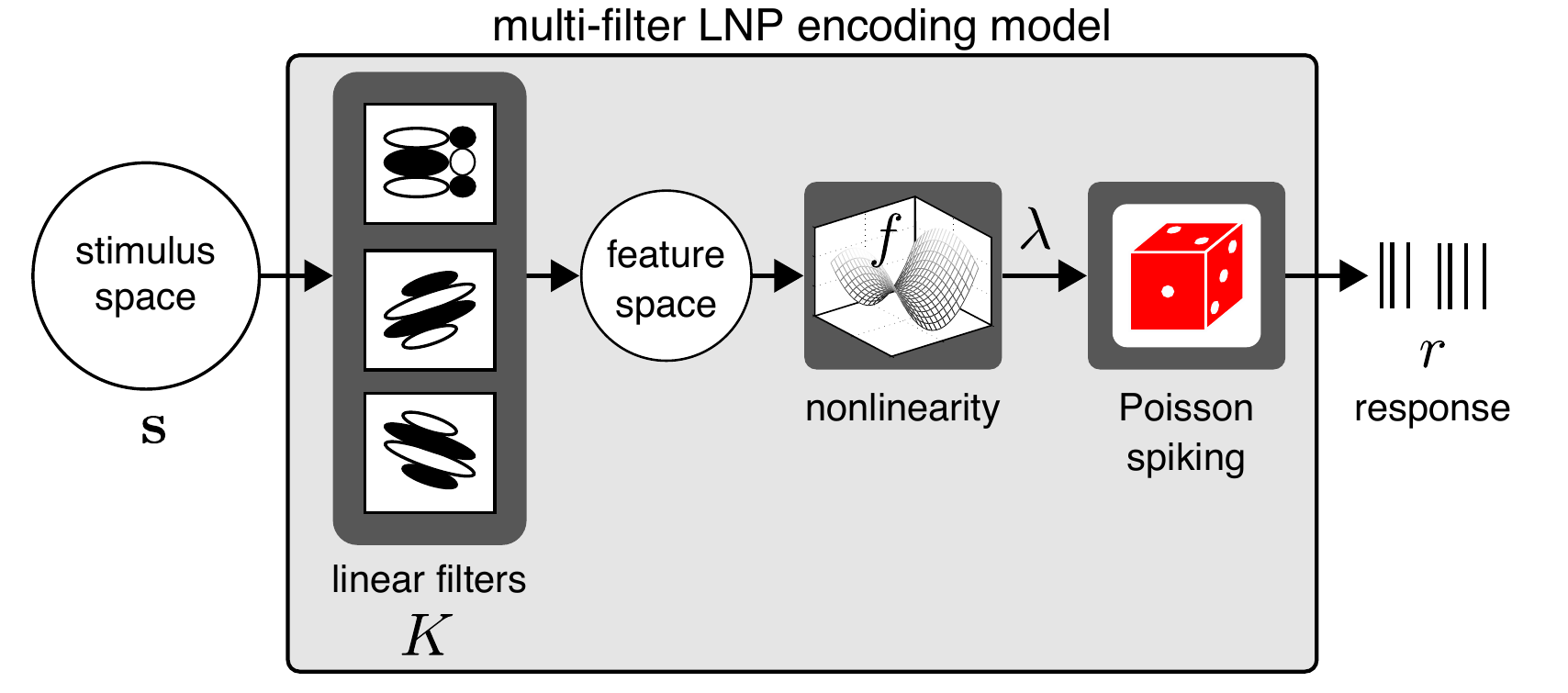}
  \end{center}
  \caption{The linear-nonlinear-Poisson (LNP) encoding model
    formalizes the neural encoding process in terms of a cascade of
    three stages. First, the high-dimensional stimulus $\vs$ projects
    onto bank of filters contained in the columns of a matrix $K$,
    resulting in a point in a low-dimensional neural feature space
    $K\trp\vs$.  Second, an instantaneous nonlinear function $f$ maps
    the filtered stimulus to an instantaneous spike rate $\lambda$.
    Third, spikes $r$ are generated according to an inhomogeneous
    Poisson process.
}
  \label{fig:intro}
\end{figure}

Linear-nonlinear cascade models provides a useful framework for
describing neural responses to high-dimensional stimuli.  These models
define the response in terms of a cascade of linear, nonlinear, and
probabilistic spiking stages (see \figref{intro}). The linear stage
reduces the dimensionality by projecting the high-dimensional stimulus
onto onto a set of linear filters, and a nonlinear function then
converts the outputs of these filters to a non-negative spike rate.

Let $\theta=\{K,\alpha\}$ denote the parameters of the LNP model,
where $K$ is a (tall, skinny) matrix whose columns contain the
stimulus filters (for cases with a single filter, we will denote the
filter with a vector $\vk$ instead of the matrix $K$), and $\alpha$ are parameters governing the nonlinear
function $f$ from feature space to instantaneous spike rate.  Under
this model, the probability of a spike response $r$ given stimulus
$\vs$ is governed by a Poisson distribution:
\begin{eqnarray}
  \lambda &=&  f(K\trp\vs)  \nonumber \\
  p(r|\lambda) &=& \tfrac{1}{r!} (\Delta\lambda)^r
  e^{-\Delta\lambda},
\label{eq:lnp}
\end{eqnarray}
where $\lambda$ denotes the stimulus-driven spike rate (or
``conditional intensity'') and $\Delta$ denotes a (finite) time bin
size.  The defining feature of a Poisson process is that responses in
non-overlapping time bins are conditionally independent given the
spike rate.  In a discrete-time LNP model, the conditional probability
of a dataset $D= \{(\vs_t,r_t)\}$, consisting of stimulus-response
pairs indexed by $t\in\{1,\ldots,N\}$, is the product of independent
terms.  The log-likelihood is therefore a sum over time bins:
\begin{equation}
  \label{eq:logli}
  \LL_{lnp} (\theta;D) \;=\;
  \sum_{t=1}^N \log  p(r_t|\vs_t,\theta) 
  \;=\; \sum_{t=1}^N \Big (r_t \log (\Delta f(K\trp\vs_t)) -  \Delta f(K\trp\vs_t)
     \Big) \;-\;  \Big(\sum_{t=1}^N \log r_t! \Big),
\end{equation}
where $- (\sum \log r_t!) $ is a constant that does not depend
on $\theta$. The ML estimate for $\theta$ is simply the maximizer of
the log-likelihood:
$ \widehat{\theta}_{ML} = \argmax{\theta} \LL_{lnp}(\theta;D)$.

\subsubsection*{Maximally informative dimensions (MID)}

The maximally informative dimensions (MID) estimator seeks to find an
informative low-dimensional projection of the stimulus by maximizing 
an information-theoretic quantity known as the 
single-spike information \cite{Brenner00b}. This quantity, which we
denote $\Iss$, is the the average information that the time of a
single spike (considered independently of other spikes) carries about
the stimulus

Although first introduced as a quantity that can be computed from the
peri-stimulus time histogram (PSTH) measured in response to a repeated
stimulus, the single-spike information 
can also be expressed as the Kullback-Leibler (KL) divergence between
two distributions over the stimulus (see \cite{Brenner00b}, appendix
B):
  \begin{equation}
  \label{eq:Isp}
  \Iss = \int \pcondsp \log
  \frac{\pcondsp}{\pstim} d\vs = 
  \kl{\pcondsp}{\pstim},
\end{equation}
where $\pstim$ denotes the marginal or ``raw'' distribution over
stimuli, and $\pcondsp$ is the distribution over stimuli conditioned
on observing a spike, also known as the ``spike-triggered'' stimulus
distribution. Note that $\pcondsp$ is {\it not} the same as
$p(\vs|r=1)$, the distribution of stimuli conditioned on a spike count
of $r=1$, since a stimulus that elicits two spikes will contribute
twice as much to the spike-triggered distribution as a stimulus that
elicits only one spike.

\begin{figure}[!t]
  \begin{center}
    \includegraphics[width=0.95\columnwidth]{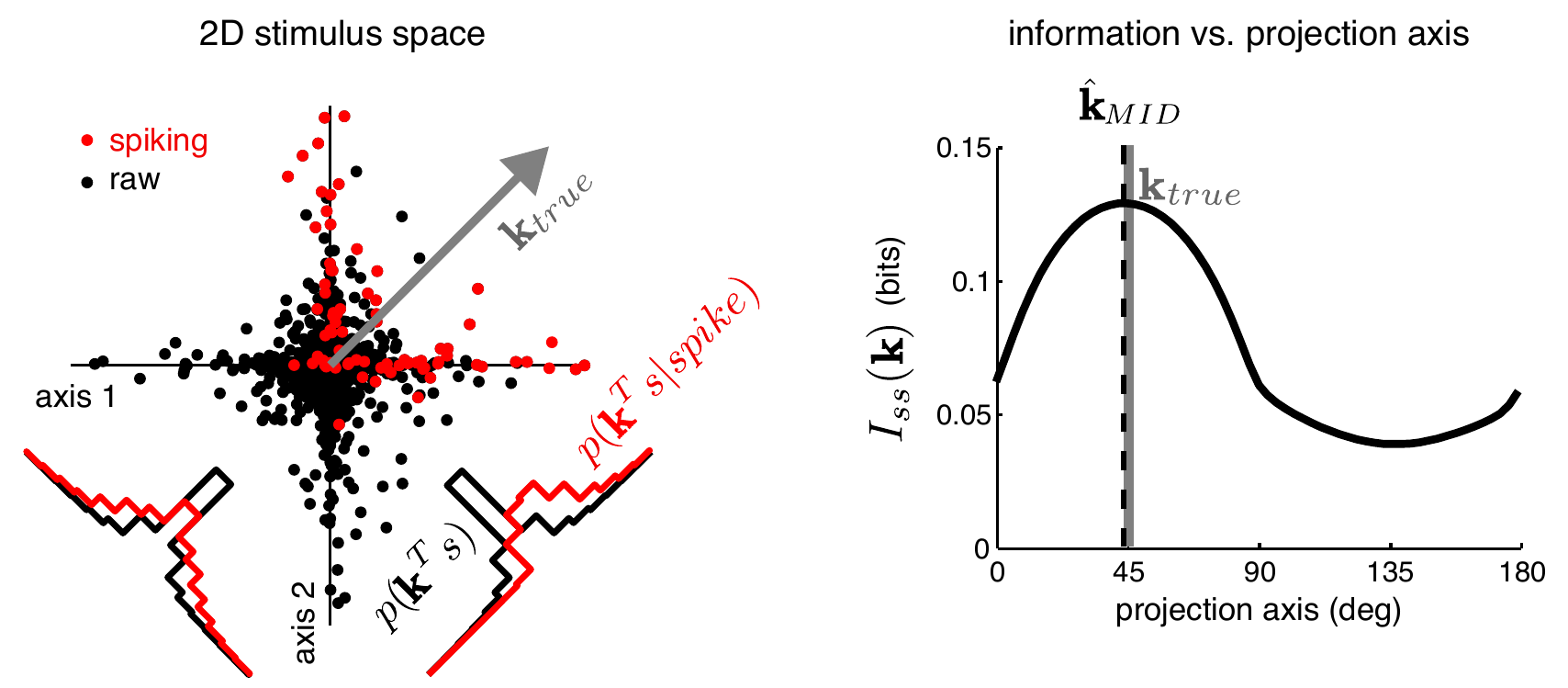}
  \end{center}
  \caption{Geometric illustration of maximally-informative-dimensions
    (MID). \textbf{Left:} A two-dimensional stimulus space, with
    points indicating the location of raw stimuli (black) and
    spike-eliciting stimuli (red).  For this simulated example, the
    probability of spiking depended only on the projection onto a
    filter $\vk_{true}$, oriented at 45$^\circ$. Histograms (inset) show
    the one-dimensional distributions of raw (black) and spike-triggered
    stimuli (red) projected onto $\vk_{true}$ (lower right) and its
    orthogonal complement (lower left).  \textbf{Right:} Estimated
    single-spike information captured by a 1D subspace, as a function
    of the axis of projection.  The MID estimate $\hat{\vk}_{MID}$ (dotted)
    corresponds to the axis maximizing single-spike information, which
    converges asymptotically to $\vk_{true}$ with dataset size.}
  \label{fig:mid}
\end{figure}

The MID estimator \cite{Sharpee04} seeks to find the linear projection
that preserves maximal single-spike information:
\begin{equation} \label{eq:Iss} 
\Iss(K) =
  \kl{p(K\trp\vs|spike)}{p(K\trp\vs)},
\end{equation}
where $p(K\trp\vs)$ and $p(K\trp\vs|spike)$ are the raw and
spike-triggered stimulus distributions projected onto the subspace
defined by the columns of $K$, respectively.  In practice, the MID
estimator maximizes an estimate of the projected single-spike
information:
\begin{equation}
  \Kmid = \arg \max_K\;  \Issempir(K), 
\end{equation}
where $\Issempir(K)$ denotes an empirical estimate of $\Iss(K)$.  The
columns of $\Kmid$ can be conceived as ``directions'' or ``axes'' in
stimulus space that are most informative about a neuron's probability
of spiking, as quantified by single-spike information. \figref{mid} 
shows a simulated example illustrating the MID estimate for a single
linear filter in a two-dimensional stimulus space.

\subsection*{Equivalence of MID and maximum-likelihood LNP}
\label{sec:mid_is_exactly_equivalent_to_maximum_likelihood_estimation_of_the_lnp_model}

Previous work has shown that MID converges asymptotically to the
maximum-likelihood (ML) estimator for an LNP model in the limit of
small time bins \cite{Kouh09,Rajan13}.  Here we present a stronger
result, showing that the equivalence is not merely asymptotic. We show
that standard MID, using histogram-based estimators for raw and
spike-triggered stimulus densities $\pstim$ and $\pcondsp$, is exactly
the ML estimator for the parameters of an LNP model, regardless of
spike rate, the time bins used to count spikes, or the amount of data.

The standard implementation of MID \cite{Sharpee04,Kouh09} uses
histograms to estimate the projected stimulus densities $p(K\trp\vs)$
and $p(K\trp\vs|spike)$. These density estimates are then used to
compute $\Issempir(K)$, the plugin estimate of single-spike
information in a subspace defined by $K$ (\eqref{Iss}).  We will now
unpack the details of this estimate in order to show its relationship
to the LNP model log-likelihood.

Let $\{B_1,\ldots,B_m\}$ denote a group of sets (``histogram bins'')
that partition the range of the projected stimuli $K\trp\vs$. In the
one-dimensional case, we typically choose these sets to be intervals
$B_i = [b_{i-1},b_i)$, defined by bin edges $\{b_0, \ldots, b_m\}$,
where $b_0=-\infty$ and $b_m = +\infty$. Then let $\vphat =
(\phat_1,\ldots, \phat_m)$ and $\vqhat = (\qhat_1,\ldots \qhat_m)$
denote histogram-based estimates of $p(K\trp\vs)$ and
$p(K\trp\vs|spike)$, respectively, given by:
\begin{equation}\begin{split}
\phat_i \quad &= \quad \frac{\textrm{\# stimuli in $B_i$}}{\textrm{\# stimuli}} \quad = \quad  \frac{1}{N}
 \sum_{t=1}^N \indic{B_i}(\vx_t)
 \\
  \qhat_i    \quad &= \quad \frac{\textrm{\# stimuli in
      $B_i$ $|$ $spike$}}{\textrm{\# $spikes$}} \quad = \quad
  \frac{1}{\nsp} \sum_{t=1}^N \indic{B_i}(\vx_t) r_t,
\label{eq:histLNP}
\end{split}\end{equation}
where $\vx_t = K\trp \vs_t$ denotes the linear projection of the
stimulus $\vs_t$, $\nsp=\sum_{t=1}^N r_t$ is the total number of
spikes, and $\indic{B_i}(\cdot)$ is the indicator function for the set
$B_i$, defined as:
\begin{equation} \label{eq:indicator}
  \indic{B_i}(\vx) = \left\{ \begin{array}{cl} 1, & \vx \in B_i \\ 0, & \vx
      \notin B_i
    \end{array} \right.
\end{equation}
The estimates $\vphat$ and $\vqhat$ are also known as the ``plug-in''
estimates, and correspond to maximum likelihood estimates for the
densities in question. These estimates give us a plug-in estimate for
projected single-spike information:
\begin{equation}
  \Issempir \;=\; \sum_{i=1}^m \qhat_i \log \frac{\qhat_i}{\phat_i} 
  \;=\; \frac{1}{\nsp} \sum_{i=1}^m \sum_{t=1}^N \indic{B_i}(\vx_t) r_t \log
  \frac{\qhat_i}{\phat_i} \;
  = \; \frac{1}{\nsp} \sum_{t=1}^N r_t \log \ghat(\vx_t)
  \label{eq:Issempir}
\end{equation}
where the function $\ghat(\vx)$ denotes the ratio of density
estimates:
\begin{equation}
 \label{eq:ghat}
\ghat(\vx) \triangleq \sum_{i=1}^m \indic{B_i}(\vx)
  \frac{\qhat_{i}}{\phat_{i}}.
%
\end{equation}
Note that $\ghat(\vx)$ is a piece-wise constant function that takes the
value $\qhat_i/\phat_i$ over the $i$th histogram bin $B_i$.

Now, consider an LNP model in which the nonlinearity $f$ is
parametrized as a piece-wise constant function, taking the value $f_i$
over histogram bin $B_i$.  Given a projection matrix $K$, the ML
estimate for the parameter vector $\alpha = (f_1, \ldots, f_m)$ is the
average number of spikes per stimulus in each histogram bin, divided
by time bin width $\Delta$, that is:
\begin{equation} \label{eq:fnlin}
  \fhat_i = \frac{1}{\Delta}\cdot \frac{\sum_{t=1}^N \indic{B_i}(\vx_t) r_t}{\sum_{t=1}^N
    \indic{B_i}(\vx_t)} = \left(\!\tfrac{\nsp}{N\Delta}\!\right) \frac{\qhat_i}{\phat_i}
\end{equation}
Note that functions $\fhat$ and $\ghat$ are related by $\fhat(\vx) =
\left(\frac{\nsp}{N\Delta}\right) \ghat(\vx)$ and that the sum
$\sum_{t=1}^N \fhat(\vx_t) \Delta = \nsp$.  We can therefore rewrite
the LNP model log-likelihood (\eqref{logli}):
\begin{eqnarray}
  \LL_{lnp}(\theta;D) &=&  \sum_{t=1}^N  r_t \log \left(
    \frac{\nsp}{N}\ghat(\vx_t)\right) \;-\; \nsp  - \sum_{t=1}^N \log r_t!
  \nonumber \\
  &=& \sum_{t=1}^N r_t \log \ghat(\vx_t) \;+\; \nsp \left(\log
    \frac{\nsp}{N}  - 1\right) - \sum_{t=1}^N \log r_t!
\end{eqnarray}
This allows us to directly relate the empirical single-spike
information (\eqref{Issempir}) with the LNP model log-likelihood,
normalized by the spike count as follows:
\begin{eqnarray}
  \Issempir(K) &=& \tfrac{1}{\nsp} \LL_{lnp}( \theta;D) \;-\;
  \tfrac{1}{\nsp}\left[ \nsp \log
    \frac{\nsp}{N}  - \nsp - \sum \log r_t!
  \right].   \\
  &=& \tfrac{1}{\nsp} \LL_{lnp}(\theta;D) \;-\;
  \tfrac{1}{\nsp}\LL_{lnp}(\theta_0,D) 
 \label{eq:logliequiv}
\end{eqnarray}
where $\LL_{lnp}(\theta_0,D)$ denotes the Poisson log-likelihood under
a ``null'' model in which spike rate does not depend on the stimulus,
but takes constant rate $\lambda_0 = \frac{\nsp}{N\Delta}$ across the
entire stimulus space.  In fact, the quantity $-\LL_{lnp}(\theta_0,D)$
can be considered an estimate for the marginal entropy of the response
distribution, $H(r) = -\sum p(r) \log p(r)$, since it is the average
log-probability of the response under a Poisson model, independent of
the stimulus. This makes it clear that the single-spike information
$\Iss$ can be equally regarded as ``LNP information''.

Empirical single-spike information is therefore equal to LNP model
log-likelihood per spike, plus a constant that does not depend on
model parameters. This equality holds independent of time bin size
$\Delta$, the number of samples $N$ and the number of spikes
$\nsp$. From this relationship, it is clear that the linear projection
$K$ that maximizes $\Issempir$ also maximizes the LNP log-likelihood
$\LL_{lnp}(\theta;D)$, meaning that the MID estimate is precisely the
same as an ML estimate for the filters in an LNP model:
\begin{equation}
\widehat K_{MID} =  \widehat{K}_{ML}.
\end{equation}
Moreover, the histogram-based estimates of the raw and spike-triggered
stimulus densities $\vphat$ and $\vqhat$, which are used for computing
the empirical single-spike information $\Issempir$, correspond to a
particular parametrization of the LNP model nonlinearity $f$ as a
piece-wise constant function over histogram bins. The ratio of these
plug-in estimates give rise to the ML estimate for $f$.  MID is thus
formally equivalent to an ML estimator for both the linear filters and
the nonlinearity of an LNP model.



\begin{figure}[!t] \begin{center}
\includegraphics[width=1\columnwidth]{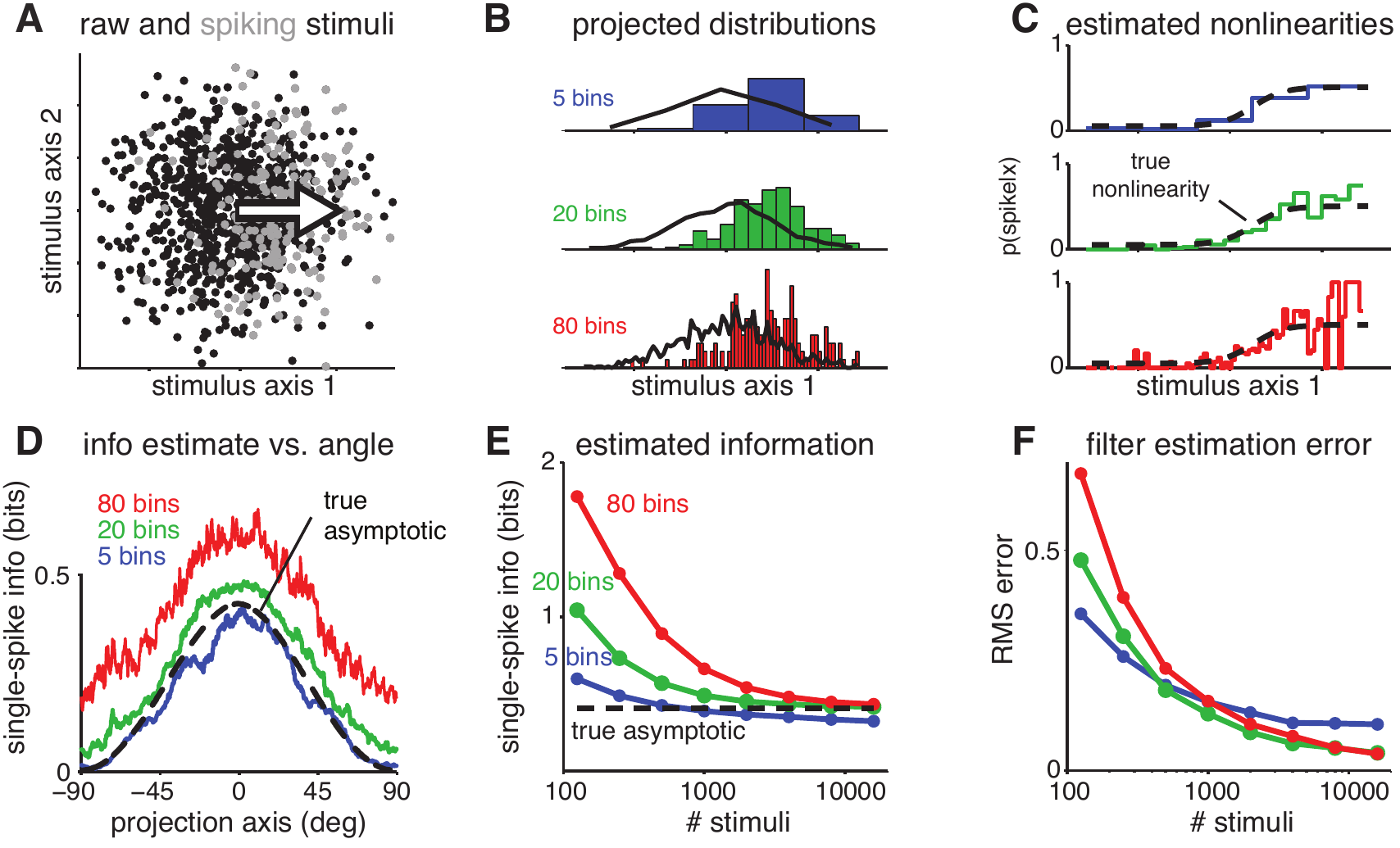}
\caption{Effects of the number of histogram bins on empirical
  single-spike information and MID performance.  \textbf{(A)} Scatter
  plot of raw stimuli (black) and spike-triggered stimuli (gray) from
  a simulated experiment using two-dimensional stimuli to drive a
  linear-nonlinear-Bernoulli neuron with sigmoidal nonlinearity.
  Arrow indicates the direction of the true filter $\vk$.
  \textbf{(B)} Plug-In estimates of $p(\vk\trp\vs|spike)$, the
  spike-triggered stimulus distribution along the true filter axis,
  from 1000 stimuli and 200 spikes, using 5 (blue), 20 (green) or 80
  (red) histogram bins.  Black traces show estimates of raw
  distribution $p(\vk\trp\vs)$ along the same axis.  \textbf{(C)} True
  nonlinearity (black) and ML estimates of the nonlinearity (derived
  from the ratio of the density estimates shown in B).  Roughness of
  the 80-bin estimate (red) arises from undersampling, or
  (equivalently) overfitting of the nonlinearity.  \textbf{(D)}
  Empirical single-spike information vs.\ direction, calculated using
  5, 20 or 80 histogram bins.  Note that the 80-bin model
  overestimates the true asymptotic single-spike information at the
  peak by a factor of more than $1.5$.  \textbf{(E)} Convergence of
  empirical single-spike information along the true filter axis as a
  function of sample size.  With small amounts of data, all three
  models overfit, leading to upward bias in estimated information.
  For large amounts of data, the 5-bin model underfits and therefore
  under-estimates information, since it lacks the smoothness to
  adequately describe the shape of the sigmoidal nonlinearity.
  \textbf{(F)} Filter error as a function of the number of stimuli,
  showing that the optimal number of histogram bins depends on the
  amount of data.}
\label{fig:Issvsnbins} 
\end{center} 
\end{figure}

Previous literature has not emphasized that the MID estimator
implicitly provides an estimate of the LNP model nonlinearity, or that
the number of histogram bins corresponds to the number of parameters
governing the nonlinearity.  Selecting the number of parameters for
the nonlinearity is important both for accurately estimating
single-spike information from finite data and for successfully finding
the most informative filter or filters. Fig. \ref{fig:Issvsnbins}
illustrates this point using data from a simulated neuron with a
single filter in a two-dimensional stimulus space.  For small datasets, the MID
estimate computed with many histogram bins (i.e., many parameters for
the nonlinearity) substantially overestimates the true $\Iss$ and
yields large errors in the filter estimate $\widehat K$.  Even with
1000 stimuli and 200 spikes, a 20-bin histogram gives substantial
upward bias in the estimate of single-spike information
(Fig.~\ref{fig:Issvsnbins}D).  Parametrization of the nonlinearity is
therefore an important problem that should be addressed explicitly
when using MID, e.g., by cross-validation or other model selection
methods.

\subsection*{Models with Bernoulli spiking}

Under the discrete-time inhomogeneous Poisson model considered above,
spikes are modeled as conditionally independent given the stimulus,
and the spike count in a discrete time bin has a Poisson distribution.
However, real spike trains may exhibit more or less variability than a
Poisson process \cite{Maimon09}. In particular, the Poisson assumption
breaks down when the time bin in which the data are analyzed
approaches the length of the refractory period, since in that case
each bin can contain at most one spike. In that case, a Bernoulli
model provides a more accurate description of neural data, since it
assigns allows only 0 or 1 spike per bin.  In fact, the Bernoulli and
discrete-time Poisson models approach the same limiting Poisson
process as the bin size (and single-bin spike probability) approaches
zero while the average spike rate remains constant.  However, as long
as single-bin spike probabilities remain measurable, the two models
differ.

Here we show that the standard ``Poisson'' MID estimator does not
necessarily maximize information between stimulus and response when
spiking is non-Poisson.  That is, if the spike count $r$ given
stimulus $\vs$ is not a Poisson random variable, then MID does not
necessarily find the subspace preserving maximal information between
stimulus and response.  To show this, we derive the mutual information
between the stimulus and a Bernoulli distributed spike count, and show
that this quantity is closely related to the log-likelihood under a
linear-nonlinear-Bernoulli encoding model.

\subsubsection*{Linear-nonlinear-Bernoulli (LNB) model}

We can define the Linear-nonlinear-Bernoulli (LNB) model by analogy to
the LNP model, but with Bernoulli instead of Poisson spiking.  The
parameters $\theta=\{K,\alpha\}$ consist of a matrix $K$ that
determines a linear projection of the stimulus space, and a set of
parameters $\alpha$ that govern the nonlinearity $f$.  Here, the
output of $f$ is spike probability $\lambda$ in the range $[0,1]$.
The probability of a spike response $r\in\{0,1\}$ given stimulus $\vs$
is governed by a Bernoulli distribution.  We can express this
model as
\begin{eqnarray}
  \lambda &=&  f(K\trp\vs)  \\
  p(r|\lambda) &=& \lambda^r (1-\lambda)^{1-r},
\label{eq:lnb}
\end{eqnarray}
and the log-likelihood for a dataset $D = \{(\vs_t,r_t)\}$ is
\begin{equation}
   \label{eq:logliBer}
   \LL_{lnb} (\theta;D)   \;=\; \sum_{t=1}^N \Big (r_t \log
   (f(K\trp\vs_t)) - (1-r_t) \log (1-f(K\trp\vs_t))
   \Big).
\end{equation}
If $K$ has a single filter and the nonlinearity is restricted to be a
logistic function, $f(x) = 1/(1+\exp(-x))$, this reduces to the
logistic regression model.  Note that the spike probability $\lambda$
is analogous to the single-bin Poisson rate $\lambda \Delta$ from the
LNP model (\eqref{lnp}), and the two models become identical in the
small-bin limit where the probability of spiking $p(r=1)$ goes to zero
\cite{Brenner00b,Rajan13}.

\subsubsection*{Bernoulli information}

We can derive an equivalent dimensionality-reduction estimator in
information-theoretic terms. The mutual information between the
projected stimulus $\vx = K\trp\vs$ and a Bernoulli spike response
$r\in\{0,1\}$ is given by:
\begin{align}
  I(\vx, r) 
    &=\; H(\vx) - H(\vx | r) \nonumber
\\
    &=\; - \int d\vx\, p(\vx) \log p(\vx) 
   \;+\; \sum_{j\in\{0,1\}}\; p(r=j) \int d\vx\,  p(\vx | r=j) \log p(\vx |
   r = j) \nonumber \\
    &=\;  \sum_{j\in\{0,1\}}\; p(r=j) \int d\vx\, p(\vx | r=j) \log\frac{p(\vx | r =
      j)}{p(\vx)} \nonumber \\
    &=\;  \sum_{j\in\{0,1\}}\; p(r=j)  \kl{p(\vx |
      r = j)}{p(\vx)}\,.
\label{eq:MIber}
\end{align}
If we normalize by the probability of observing a spike, we obtain a
quantity with units of bits-per-spike that can be directly compared to
single-spike information. We refer to this as the Bernoulli
  information:
\begin{equation} \label{eq:Iber} 
\Iber = \frac{1}{p(r=1)} I(\vx,r)
  = I_0 + I_{ss}
\end{equation}
where $I_0 = \frac{p(r=0)}{p(r=1)} \kl{p(\vx|r=0)}{p(\vx)}$ is the
information (per spike) carried by silences and $\Iss$ is the
single-spike information (eq.~\ref{eq:Iss}). 
Thus, where single-spike information quantifies the information
conveyed by each spike alone (no matter how many spikes might co-occur
in the same time bin) but neglects the information conveyed by the
absence of any spike, the Bernoulli information quantifies information
per bin, whether a (by assumption, single) spike appears within it or
not.

Let $\Iberempir = \widehat I_0 + \Issempir$ denote the empirical or
plug-in estimate of the Bernoulli information, where $\Issempir$ is the
empirical single-spike information (\eqref{Issempir}), and $\widehat
I_0$ is a plug-in estimate of the KL divergence between $p(\vx|r=0)$
and $p(\vx)$, weighted by $(N-\nsp)/\nsp$, the ratio of the number of
silences to the number of spikes. It is straightforward to show that
empirical Bernoulli information equals the LNB model log-likelihood per
spike plus a constant:
\begin{equation} \label{eq:Iberequiv}
  \Iberempir = \frac{1}{\nsp} \LL_{lnb} + \frac{1}{\bar r} \widehat H[r]
\end{equation}
where $\bar r= \frac{\nsp}{N}$ denotes mean spike count per bin and
$\widehat H[r] = - \frac{\nsp}{N} \log \frac{\nsp}{N} -
\frac{N-\nsp}{N} \log \frac{N-\nsp}{N} $ is the plug-in estimate for
the marginal response entropy.  
Because the second term
is independent of $\theta$, the maximum of the empirical Bernoulli
information is identical to the maximum of the LNB model likelihood,
meaning that once again, we have an exact equivalence between
likelihood-based and information-based estimators.

\subsubsection*{Failure modes for MID under Bernoulli spiking}

The empirical Bernoulli information is strictly greater than the
estimated single-spike (or ``Poisson'') information for a binary spike
train that is not all zeros or ones, since $\widehat I_0>0$ and these
spike absences are neglected by the single-spike information measure.
Only in the limit of infinitesimal time bins, where $p(r=1)
\rightarrow 0$, 
does $\Iberempir$ converge to $\Issempir$ \cite{Brenner00b,Rajan13}.
As a result, standard MID can fail to identify the most informative
subspace when applied to a neuron with Bernoulli spiking.  We
illustrate this phenomenon with two (admittedly toy) simulated
examples. For both examples, we compute the standard MID estimate
$\widehat \vk_{MID}$ by maximizing $\Issempir$, and the LNB filter
estimate $\widehat \vk_{Ber}$ which maximizes the LNB likelihood, or
equivalently $\Iberempir = \Issempir + \widehat{I}_0$.

\begin{figure}[!t]
  \begin{center}
    \includegraphics[width=0.9\columnwidth]{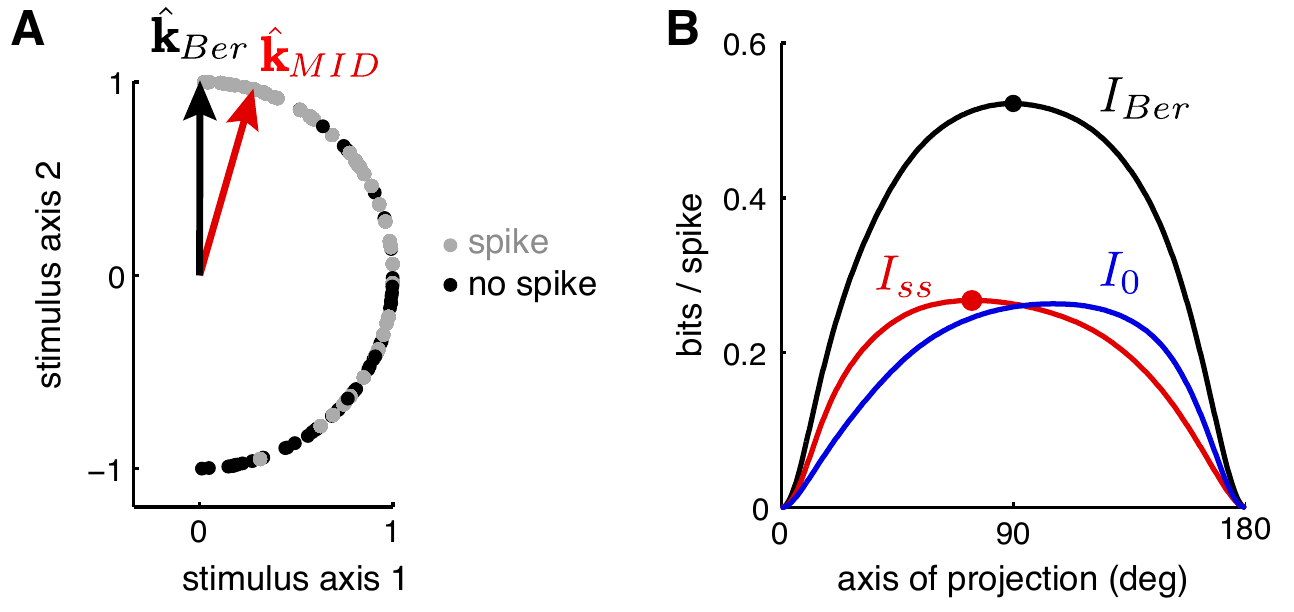} 
    \caption{Illustration of MID failure mode due to non-Poisson
      spiking.  {\bf (A)} Stimuli were drawn uniformly on the unit
      half-circle, $\theta \sim \textrm{Unif}(-\pi/2,\pi/2)$. The
      simulated neuron had Bernoulli (i.e., binary) spiking, where the
      probability of a spike increased linearly from 0 to 1 as
      $\theta$ varied from -$\pi/2$ to $\pi/2$, that is:
      $p(spike|\theta) = \theta/\pi + 1/2$.  Stimuli eliciting
      ``spike'' and ``no-spike'' are indicated by gray and black
      circles, respectively.  For this neuron, the most informative one-dimensional
      linear projection corresponds to the vertical axis ($\widehat
      \vk_{Ber}$), but the MID estimator ($\widehat \vk_{MID}$) exhibits a
      $16^\circ$ clockwise bias.  {\bf (B)} Information from spikes
      (black), silences (gray), and both (red), as a function of
      projection angle. The peak of the Bernoulli information (which
      defines $\widehat \vk_{Ber}$) lies close to $\pi/2$, while the
      peak of single-spike information (which defines $\widehat
      \vk_{MID}$) exhibits the clockwise bias shown in A.  Note that
      $\widehat \vk_{MID}$ does not converge to the optimal direction
      even in the limit of infinite data, due to its lack of
      sensitivity to information from silences. Although this figure
      is framed in an information-theoretic sense, equations (19) and
      (20) detail the equivalence between $I_{Ber}$ and
      $\mathcal{L}_{lnb}$, so that this figure can be viewed from
      either an information-theoretic or likelihood-based
      perspective.}
       \label{fig:Counter2}
  \end{center}
\end{figure}

The first example (Fig.~\ref{fig:Counter2}) uses raw stimuli uniformly
distributed on the right half of the unit circle. The Bernoulli spike
probability $\lambda$ increases linearly as a function of stimulus
angle: $\lambda = (\vs-\pi/2)/\pi$, for $\vs \in (-\pi/2, \pi/2]$.
For this neuron, the most informative 1D axis is the vertical axis,
which is closely matched by the estimate $\widehat \vk_{Ber}$. By contrast,
$\widehat \vk_{MID}$ exhibits a substantial clockwise bias, resulting from
its failure to take into account the information from silences (which
are more informative when spike rate is
high). Fig.~\ref{fig:Counter2}B shows the breakdown of total Bernoulli
information into $\Issempir$ (spikes) and $\widehat I_0$ (silences) as a
function of projection angle, which illustrates the relative biases of
the two quantities.

\begin{figure}[!t]
 \begin{center}
   \includegraphics[width=1\columnwidth]{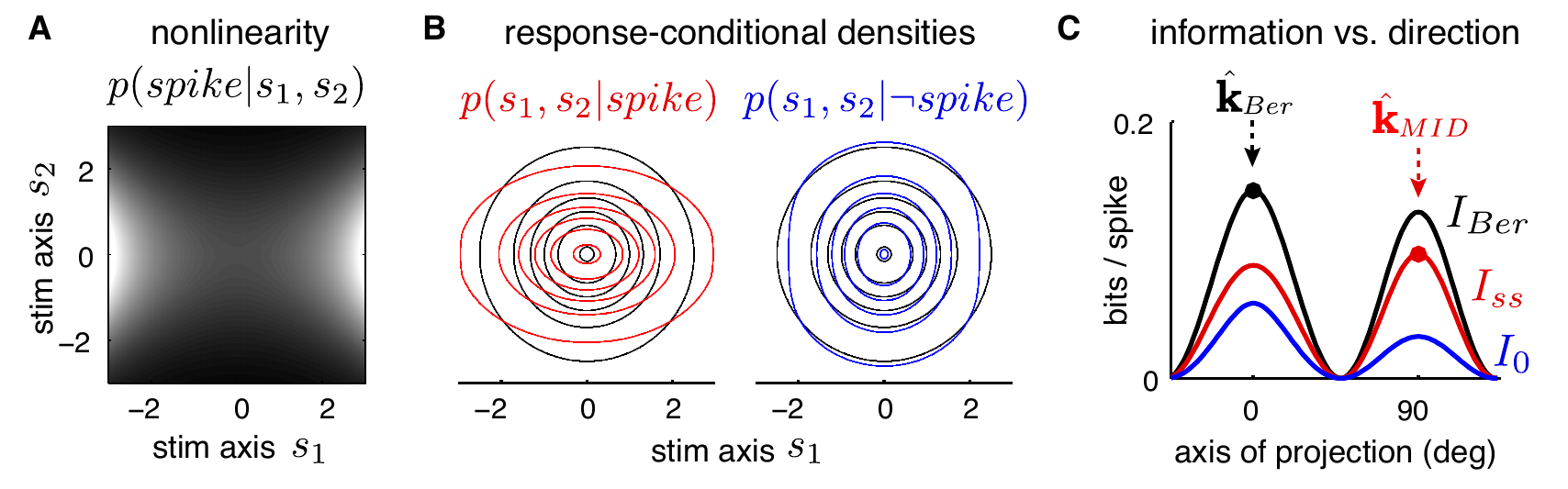} 
   \caption{A second example Bernoulli neuron for which $\widehat
     \vk_{MID}$ fails to identify the most-informative one-dimensional subspace. The
     stimulus space has two dimensions, denoted $s_1$ and $s_2$, and
     stimuli were drawn $iid$ from a standard Gaussian $\Nrm(0,1)$.
     \textbf{(A)} The nonlinearity $f(s_1,s_2)=p(spike|s_1,s_2)$ is
     excitatory in $s_1$ and suppressive in $s_2$; brighter intensity
     indicates higher spike probability.  \textbf{(B)} Contour plot of
     the stimulus-conditional densities given the two possible
     responses: ``spike" (red) or ``no-spike'' (blue), along with the
     raw stimulus distribution (black).  \textbf{(C)} Information
     carried by silences ($I_0$), single spikes ($\Iss$), and total
     Bernoulli information ($\Iber = I_0 + \Iss$) as a function of
     subspace orientation.  The MID estimate $\widehat
     \vk_{MID}=90^\circ$ is the maximum of $\Iss$, but the total
     Bernoulli information is in fact 13\% higher at $\widehat
     \vk_{Ber}=0^\circ$ due to the incorporation of no-spike
     information. Although both stimulus axes are clearly relevant to
     the neuron, MID identifies the less informative one. As with the
     previous figure, equations (19) and (20) detail the equivalence
     between $I_{Ber}$ and $\mathcal{L}_{lnb}$, so that this figure can be
     viewed from either an information-theoretic or
     likelihood-based perspective.}
   \label{fig:Counter}
 \end{center}
\end{figure}

A second example (Fig.~\ref{fig:Counter}) uses stimuli drawn from a
standard bivariate Gaussian (0 mean and identity covariance), in which
standard MID makes a $\pi/2$ error in identifying the most informative
one-dimensional subspace.  The neuron's nonlinearity (Fig.~\ref{fig:Counter}A) is
excitatory in stimulus axis $s_1$ and suppressive in stimulus axis
$s_2$ (indicating that a large projection onto $s_1$ increases spike
probability, while a large projection onto $s_2$ decreases spike
probability).  For this neuron, both stimulus axes are clearly
informative, but the (suppressive, vertical) axis $s_2$ carries 13\%
more information than the (excitatory, horizontal) axis $s_1$.
However, the standard MID estimator identifies $s_1$ as the most
informative axis (Fig. \ref{fig:Counter}C), due once again to the
failure to account for the information carried by silences.

These artificial examples were designed to emphasize the information
carried by missing spikes, and we do not expect such stark differences
between Bernoulli and Poisson estimators to arise in the general case
of neural data.  However, it is clear that the assumption of Poisson
firing can lead the standard MID estimator to make mistakes when
spiking is actually Bernoulli (or generated by some other
distribution).  In general, we suggest that the question of which
estimator performs better is an empirical one, and depends on which
model (Bernoulli or Poisson) describes the true spiking process more
accurately.

\subsubsection*{Quantifying MID information loss for binary spike trains}

In the limit of infinitesimal time bins, the information carried by
silences goes to zero, and the plug-in estimates for Bernoulli and
single-spike (``Poisson'') information converge: $I_0 \rightarrow 0$
and $\Iberempir \rightarrow \Issempir$.  However, for finite time
bins, the Bernoulli information can substantially exceed single-spike
information.  In the previous section, we showed that this mismatch
can lead to errors in subspace identification.  Here we derive a lower
bound on the information lost due to the neglect of $I_0$, the
information (per spike) carried by silences, as a function of marginal
probability of a spike, $p(r=1)$.

\begin{figure}[!t]
  \begin{center}
    \includegraphics[width=0.33\columnwidth]{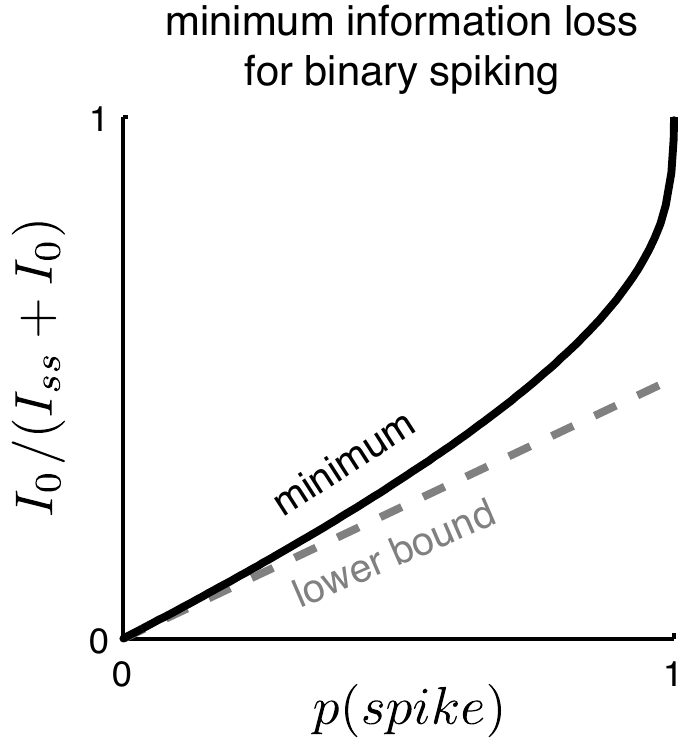}
  \end{center}
  \caption{Lower bound on the fraction of total information neglected
    by MID for a Bernoulli neuron, as a function of the marginal spike
    probability $p(spike) = p(r=1)$, for the special case of a binary
    stimulus.  Information loss is quantified as the ratio
    $I_0/(I_{0}+I_{ss})$, the information due to no-spike events,
    $I_0$, divided by the total information due to spikes and
    silences, $I_{0}+I_{ss}$.  The dashed gray line shows the lower
    bound derived in the limit $p(spike) \rightarrow 0$.
    The solid black line shows the actual minimum achieved for binary
    stimuli $s\in \{0,1\}$ with $p(s=1) = q$, computed via a numerical
    search over the parameter $q\in[0, 1]$ for each value of
    $p(spike)$.  The lower bound is substantially loose for $p(spike)
    > 0$, since as $p(spike) \rightarrow 1$, the fraction of
    information due to silences goes to 1.}
  \label{fig:bound}
\end{figure}

In the limit of rare spiking, $p(r=1) \rightarrow 0$, we find that:
\begin{equation}
\frac{I_0}{\Iber} = \frac{I_0}{I_0 + \Iss} \geq \frac{p(r=1)}{2}.
\end{equation}
The fraction of lost information is at least half the marginal spike
probability. Thus, for example, if 20\% of the bins in a binary spike
train contain a spike, the standard MID estimator will necessarily
neglect at least 10\% of the total mutual information.  We show this
bound holds in the asymptotic limit of small $p(r=1)$ (see Methods for details), but conjecture that it holds for all
$p(r=1)$. The bound is tight in the Poisson limit,
  $p(r=1)\rightarrow 0$, but is substantially loose in the limit where
  spiking is common $p(r=1)\rightarrow 1$, in which all information is
  carried by silences.  Fig.~\ref{fig:bound} shows our bound
compared to the actual (numerical) lower bound for an example with a
binary stimulus.

\subsection*{Models with arbitrary spike count distributions}

For neural responses binned at the stimulus refresh rate (e.g., 100
Hz), it is not uncommon to observe multiple spikes in a single bin.
For the general case, then, we must consider an arbitrary distribution
over counts conditioned on a stimulus.  As we will see, maximizing the
mutual information based on histogram estimators is once again
equivalent to maximizing the likelihood of an LN model with piece-wise
constant mappings from the linear stimulus projection to count
probabilities.

\subsubsection*{Linear-nonlinear-count (LNC) model}

Suppose that a neuron responds to a stimulus $\vs$ with a spike count
$r\in\{0,\ldots,\rmax\}$, where $\rmax$ is the maximum possible number
of spikes within the time bin (constrained by the refractory period or
other firing rate saturation).
The linear-nonlinear-count (LNC) model, which includes LNB as a
special case, is defined by a linear dimensionality reduction matrix
$K$ and a set of nonlinear functions $\{f{^{(0)}}\ldots,
f^{{(\rmax)}}\}$ that map the projected stimulus to the probability of
observing $\{0, \ldots, \rmax\}$ spikes, respectively.  We can write
the probability of a spike response $r$ given projected stimulus $\vx
= K\trp\vs$ as:
\begin{eqnarray}
  \lambda\jj &=&  f\jj(\vx), \quad \textrm{ for } j =0,\ldots,\rmax  \nonumber \\
  p(r=j|\lambda\jj) &=& \lambda\jj.
\label{eq:lnc}
\end{eqnarray}
Note that there is a implicit linear constraint on the functions $f$
requiring that $\sum_j f\jj(\vx) = 1, \forall \vx$, since the
probabilities over possible counts must add to 1 for each stimulus.

The LNC model log-likelihood on the parameters $\theta = (K,
\alpha^{(0)}, \ldots \alpha^{(\rmax)})$ for data $D = \{(\vs_t,r_t)\}$
can be written:
\begin{equation}
  \label{eq:logliCat}
  \LL_{lnc} (\theta;D)   
  \;=\; \sum_{t=1}^N \sum_{j=0}^{\rmax}  \Big (
  \indic{j}(r_t)\, \log \left( f\jj (K\trp\vs_t) \right)
  \Big), 
\end{equation}
where $\indic{j}(r_t)$ is an indicator function selecting time bins
$t$ in which the spike count is $j$.  As before, we consider the case
where $f\jj$ takes a constant value in each of $m$ histogram bins
$\{B_i\}$, so that the parameters are just those constant values:
$\alpha\jj = (f_0\jj, \ldots, f_{m}\jj)$. The maximum-likelihood
estimates for the values can be given in terms of the histogram
probabilities:
\begin{align}
  \fhat_i\jj \,=\, \frac{n_i\jj}{n_i} \,=\, \frac{\qhat_i\jj}{\phat_i} \frac{N\jj}{N}\,.
\end{align}
where $n_i$ is the number of stimuli in bin $B_i$, $n_i^{(j)}$ is the
number of stimuli in bin $B_i$ that elicited $j$ spikes, $N^{(j)}$ is
the number of stimuli in all bins that elicited $j$ spikes, and $N$ is
the total number of stimuli.  The histogram fractions of the projected
raw spike counts $\phat_i$ are defined as in \eqref{histLNP}, with the
$j$-spike conditioned histograms defined analogously:
\begin{align}
  \qhat_i\jj &= \frac{1}{N\jj} \sum_t \indic{B_i}(\vx_t)\,
  \indic{j}(r_t)\;=\; \frac{n_i^{(j)}}{N^{(j)}}\,,
\end{align}

Thus, the log-likelihood for projection matrix $K$, having already
maximized with respect to the nonlinearities by using their plug-in
estimates, is
\begin{align}
  \LL_{lnc} (K;D) &= \sum_{t=1}^N \sum_{j=0}^{\rmax} \Big(
  \indic{j}(r_t)\, \log \left( \fhat\jj (K\trp\vs_t) \right) \Big)
  \\
  &= \sum_{t=1}^N \sum_{j=0}^{\rmax} \sum_{i=1}^m \Big(
  \indic{j}(r_t)\,\indic{B_i}(K\trp\vs_t) \log \big( \fhat\jj_i \big)
  \Big)
  \\
  &= \sum_{j=0}^{\rmax} \sum_{i=1}^m \bigg(n_i^{(j)} \log
  \frac{n_i^{(j)}}{n_i} \bigg)
  \\
  &= \sum_{j=0}^{\rmax} \sum_{i=1}^m \bigg( N\jj \qhat_i\jj \log
  \Big( \frac{\qhat_i\jj}{\phat_i} \frac{N\jj}{N} \Big) \bigg)
  \\
  &= \sum_{j=0}^{\rmax} \sum_{i=1}^m \bigg( N\jj \qhat_i\jj \log
  \Big( \frac{\qhat_i\jj}{\phat_i} \Big) \bigg) + \sum_{j=0}^{\rmax}
  \bigg( N\jj \log \Big( \frac{N\jj}{N} \Big) \bigg)\,.
  \label{eq:logliLNCexpanded}
\end{align}

\subsubsection*{Information in spike counts}

If the binned spike-counts $r_t$ measured in response to stimuli
$\vs_t$ are not Poisson distributed, the projection matrix $K$ which
maximizes the mutual information between $K\trp\vs$ and $r$ can be
found as follows.  Recalling that $\rmax$ is the maximal spike count
possible in the time bin and writing $\vx = K\trp\vs$, we have:
\begin{align}
  I(\vx, r) 
    &= H(\vx) - H(\vx | r)
\\
    &= - \int d\vx\; p(\vx) \log p(\vx) 
       + \sum_{j = 0}^{\rmax} p(r=j) \int d\vx\; p(\vx | r=j) \log p(\vx | r = j)
\\
    &= \sum_{j = 0}^{\rmax} p(r=j) \int d\vx\; p(\vx | r=j) 
                     \log\frac{p(\vx | r = j)}{p(\vx)}
\\
    &= \sum_{j = 0}^{\rmax} p(r=j) \kl{p(\vx | r = j)}{p(\vx)}\,.
\label{eq:MIcount}
\end{align}
To ease comparison with the single-spike information, which is
measured in bits per spike, we normalize the mutual information by the
mean spike count to obtain:
\begin{equation} \label{eq:Icount}
  \Icount  =  \frac{1}{\bar r} I(\vx,r) =  I_0 + I_1 + \cdots + I_{\rmax}
\end{equation}
where $\bar r = \sum_t r_t / N$ is the mean spike count, and $I_j =
\frac{p(r=j)}{\bar r} \kl{p(\vx|r=j)}{p(\vx)}$ is the normalized
information carried by the $j$-spike responses. Note that $I_1$, the
information carried by single-spike responses, is {\it not} the same
as the single-spike information $\Iss$, since the latter combines
information from all responses with 1 or more spikes, by assuming that
each spike is conditionally independent of all other spikes.

Given experimental data $D = \{(\vs_t, r_t)\}$ the mutual information
must be estimated.  If we again use the histogram-based plug-in
estimator, we obtain:
\begin{equation} \label{eq:Icountempir}
  \Icountempir \;=\; \sum_{j=0}^{\rmax} \frac{1}{\bar r} \frac{N\jj}{N}
  \sum_{i=1}^m \qhat_i\jj \log \frac{\qhat_i\jj}{\phat_i} \,.
\end{equation}
Comparison with the LNC model log-likelihood
(\eqref{logliLNCexpanded}) reveals that:
\begin{equation}
  \Icountempir = \frac{1}{\nsp} \LL_{lnc}  +
  \frac{1}{\bar r} \widehat H[r]
\end{equation}
where $\widehat H[r] = -\sum_{j=0}^{\rmax} \frac{N\jj}{N} \log
\frac{N\jj}{N}$ is the plug-in estimate for the marginal entropy of
the observed spike counts.  Note that this also proves the
relationship between Bernoulli information and LNB model
log-likelihood (\eqref{Iberequiv}) in the special case where
$\rmax=1$.

Thus, we see that even in the general case of a completely arbitrary
distribution over spike counts given a stimulus, the subspace
projection $K$ that maximizes the histogram-based estimate of mutual
information is identical to the maximum-likelihood $K$ for an
LN model with a corresponding piece-wise constant parametrization of
the nonlinearities.

\begin{figure}[!t]
\centering
    \includegraphics[width=.99\textwidth]{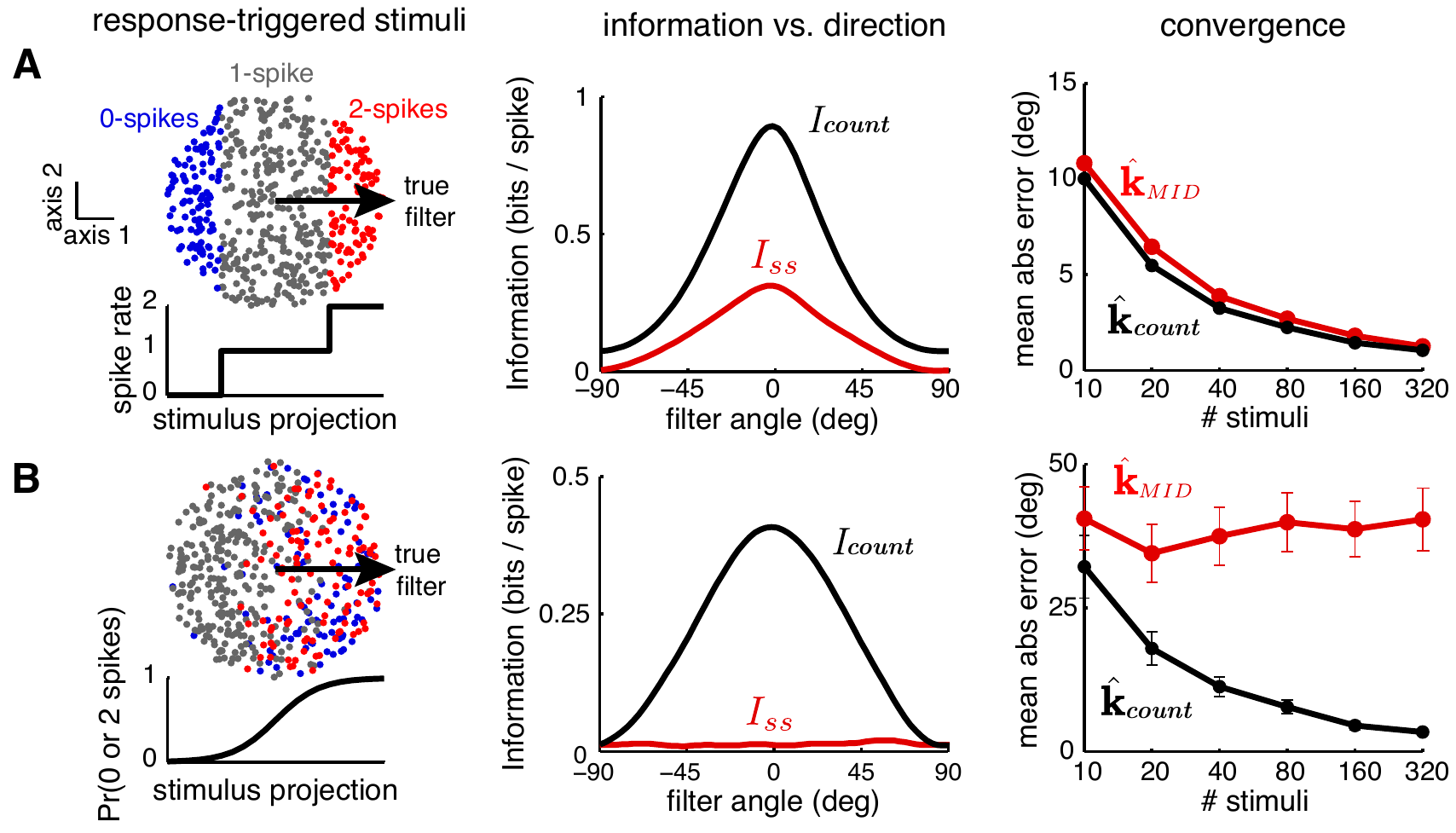}
    \caption{Two examples illustrating sub-optimality of MID under
      discrete (non-Poisson) spiking.  In both cases, stimuli were
      uniformly distributed within the unit circle and the simulated
      neuron's response depended on a 1D projection of the stimulus
      onto the horizontal axis ($\theta=0$).  Each stimulus evoked 0,
      1, or 2 spikes. 
      \textbf{(A)} Deterministic neuron.  {\it Left}: Scatter plot of
      stimuli labelled by number of spikes evoked, and the
      piece-wise constant nonlinearity governing the response (below).
      The nonlinearity sets the response count deterministically,
      thus dramatically violating Poisson expectations.
      {\it Middle}: information vs.\ axis of projection.  The total
      information $\Icount$ reflects the information from 0-, 1-, and
      2-spike responses (treated as distinct symbols), while the
      single-spike information $\Iss$ ignores silences and treats
      2-spike responses as two samples from $p({\bf s}|spike)$. 
      {\it Right}: Average absolute error in $\widehat \vk_{MID}$ and
      $\widehat \vk_{count}$ as a function of sample size; the latter
      achieves 18\% lower error due to its sensitivity to the
      non-Poisson structure of the response.
     \textbf{(B)} Stochastic neuron with sigmoidal nonlinearity
     controlling the stochasticity of responses.
     The neuron transitions from almost always emitting 1 spike for
     large negative stimulus projections, to generating either 0 or 2
     spikes with equal probability at large positive projections.
     Here, the nonlinearity does not modulate the mean spike rate, so
     $\Issempir$ is approximately zero for all stimulus projections
     (middle) and the MID estimator does not converge
     (right). However, the $\widehat \vk_{count}$ estimate converges
     because the LNC model is sensitive to the change in conditional
     response distribution. Eq. (37) details the relationship between
     $I_{count}$ and $\mathcal{L}_{lnc}$, so that this figure can be interpreted
     from either an information-theoretic or likelihood-based perspective.}
    \label{fig:discrete_examples}
\end{figure}

\subsubsection*{Failures of MID under non-Poisson count distributions}

We formulate two simple examples to illustrate the sub-optimality of
standard MID for neurons whose stimulus-conditioned count
distributions are not Poisson.  For both examples, the neuron was
sensitive to a one-dimensional projection along the horizontal axis and emitted
either 0, 1, or 2 spikes in response to a stimulus.

Both are illustrated in Fig.~\ref{fig:discrete_examples}.  The first
example (A) involves a deterministic neuron, where spike count is 0,
1, or 2 according to a piece-wise constant nonlinear function of the
projected stimulus. Here, MID does not use the information from zero
or two-spike bins optimally; it ignores information from zero-spike
responses entirely, and treats stimuli eliciting two spikes as two
independent samples from $p(\vx|spike)$.  The $\Icount$ estimator is
sensitive to the non-Poisson statistics of the response, and combines
information from all spike counts (\eqref{Icount}), yielding both
higher information and faster convergence to the true filter.

Our second example (Fig.~\ref{fig:discrete_examples}B) involves a
model neuron in which a sigmoidal nonlinearity determines the
probability that it fires exactly 1 spike (high at negative stimulus
projections) or stochastically emits either 0 or 2 spikes, each with
probability 0.5 (which becomes more probable at large positive
stimulus projections).  Thus, the nonlinearity does not change the
mean spike rate, but strongly affects its variance. Because the
probability of observing a single spike is not affected by the
stimulus, single-spike information zero for all projections, and the
MID estimate does not converge to the true filter even with infinite
data.  However, the full count information $\Icountempir$ correctly
weights the information carried by different spike counts and provides
a consistent estimator for $K$.

\subsection*{Identifying high-dimensional subspaces}
\label{sec:extending_mid_to_higher_dimensional_subspaces}

A significant drawback to standard MID is that it does not scale
tractably to high-dimensional subspaces; that is, to the simultaneous
estimation of many filters. MID has usually been limited to estimation
of only one or two filters, and we are unaware of a practical setting
in which it has been used to recover more than three.  This stands in
contrast to methods like spike-triggered covariance (STC)
\cite{deRuyter88,Schwartz06}, information-theoretic spike-triggered
average and covariance (iSTAC) \cite{Pillow06}, projection-pursuit
regression \cite{Rapela10}, Bayesian spike-triggered covariance
\cite{ParkI11}, and quadratic variants of MID
\cite{Fitzgerald11,Rajan13plos}, all of which can tractably estimate ten
or more filters.  This capability may be important, given that V1
neurons exhibit sensitivity to as many as 15 dimensions \cite{Rust05},
and many canonical neural computations (e.g., motion estimation)
require a large number of stimulus dimensions \cite{Rust06,Rajan13plos}.

Before we continue, it is helpful to consider {\it why} MID is
impractical for high-dimensional feature spaces.  The problem isn't
the number of filter parameters: these scale linearly with
dimensionality, since a $p$-filter model with $D$-dimensional stimuli
requires only $Dp$ parameters, or indeed only $(D-1)p - \frac{1}{2}
p(p-1)$ parameters to specify the subspace. The problem is instead the
number of parameters needed to specify the densities $p({\bf x})$ and
$p({\bf x}|spike)$.  For histogram-based density estimators, the number of
parameters grows exponentially with dimension: a histogram with $m$
bins along each of $p$ filter axes requires $m^p$ parameters, a
phenomenon sometimes called the ``curse of dimensionality''.

\subsubsection*{Density vs.\ nonlinearity estimation}

A key benefit of the LNP model likelihood framework is that it shifts
the focus of estimation away from the separate densities
$p(\vx|spike)$ and $p(\vx)$ to a single nonlinear function $f$.  This
change in focus makes it easier to scale the likelihood approach to
high dimensions for a few different reasons.
First, direct estimation of a single nonlinearity in place of two
densities immediately halves the number of parameters required to
achieve a similarly detailed picture of the neuron's response to the
filtered stimulus.
Second, the dependence of the MID cost function on the logarithm of
the ratio $p(\vx|spike)/p(\vx)$ makes it very sensitive to noise in
the estimated value of the denominator $p(\vx)$ when that value is
near 0.  Unfortunately, as $p(\vx)$ is also the probability with which
samples are generated, these low-value regions are precisely where the
fewest samples are available.  This is a common difficulty in the
empirical estimation of information-theoretic quantities, and others
working in more general machine-learning settings have suggested
direct estimation of the ratio rather than its parts
\cite{Sugiyama10,Sugiyama12,Suzuki13}.  In LN neural modeling such
direct estimation of the ratio is equivalent to direct estimation of
the nonlinearity.

This brings us to the third, and most subtle but perhaps most
powerful, benefit of the likelihood method's focus on $f$.
As the nonlinearity is seen to be a property of the modeled neuron
rather than of the stimulus, it may be more straightforward to
construct a valid smoothed or structured parametrization for $f$ (or
to otherwise regularize its estimate based on prior beliefs about
neuronal properties) than it is for the stimulus densities.
For example, consider an experiment using natural visual images.
While natural images presumably form a smooth manifold within the
space of all possible pixel patterns, the structure of this manifold
is neither simple nor known.  
The natural distribution of images does not factor over disjoint sets
of pixels, nor over linear projections of pixel values.  
A small random perturbation in all pixels makes a natural image appear
unnaturally noisy, violating the underlying presumption of kernel
density estimators that local perturbations do not alter the density
much.
Indeed the question of how best to model the distribution of natural
stimuli is a matter of active research.
By contrast, we might expect to be able to develop better parametric
forms to describe the non-linearities expressed by neural systems.
For instance, we might expect the neural nonlinearity to vary
smoothly in the space of photoreceptor activation, and thus of filter
outputs.  Thus, locally kernel-smoothed estimates of the non-linear
mapping---or even parametric choices of function class, such as
low-order polynomials---might be valid, even if the stimulus density
changes abruptly.
Alternatively, subunits within the neural receptive field might lead
to additively or multiplicatively separable components of the
nonlinearity that act on the outputs of different filters.  
In this case, it would be possible to factor $f$ between two subsets
of filter outputs, say to give $f(\vx) = f_1(\vx_1) f_2(\vx_2)$, even
though there is no reason for the stimulus to factor over these
filters that are defined by the neural system: $p(\vx) \neq p(\vx_1)
p(\vx_2)$.  This reduction of $f$ to two (or more) lower-dimensional
functions would avoid the exponential parameter explosion implied by
the curse of dimensionality.


Indeed, such strategies for parametrization of the nonlinear mapping
are already implicit in likelihood-based estimators inspired by the
spike-triggered average and covariance.  In many such cases, $f$ is
parametrized by a quadratic form embedded in a 1D nonlinearity
\cite{ParkI11}, so that the number of parameters scales only
quadratically with the number of filters.  A similar approach has been
formulated in information-theoretic terms using a quadratic logistic
Bernoulli model \cite{Fitzgerald11,Rajan13plos,Rajan13}.  Another method,
known as extended projection pursuit regression (ePPR) \cite{Rapela10}, has
parametrized $f$ as a sum of one-dimensional nonlinearities, in which case the
number of parameters grows only linearly with the number of filters.

\subsubsection*{Parametrizing the many-filter LNP model}

Here we provide a general formulation that encompasses both standard
MID and constrained methods that scale to high-dimensional subspaces.
We can rewrite the LNP model (\eqref{lnp}) as follows:
\begin{alignat}{3} \label{eq:manyLNP1}
  \vx &\;=\;K\trp \vs \qquad & &\textrm{(dimensionality reduction)}
  \\ \label{eq:manyLNP2}
  \lambda & \;=\; f(\vx) = g\left( \sum_{i=1}^{n_\phi} \alpha_i
    \phi_i(\vx) \right) \qquad & &
  \textrm{(nonlinearity)}    \\ \label{eq:manyLNP3}
  r|\vs \; &\sim\; \textrm{Poiss}(\lambda \Delta) \qquad & &
  \textrm{(spiking).}
\end{alignat}
The nonlinearity $f$ is parametrized using basis functions
$\{\phi_i(\cdot)\}$, $i=1,\ldots,n_\phi$, which are linearly combined
with weights $\alpha_i$ and then passed through a scalar nonlinearity
$g$.  We refer to $g$ as the output nonlinearity; its primary
role is to ensure the spike rate $\lambda$ is positive regardless of
weights $\alpha_i$. This can also be considered a special case of an LNLN
model \cite{Vintch12,Theis13,McFarland13}.

If we fix $g$ and the basis functions $\{ \phi_i\}$ in advance,
fitting the nonlinearity simply involves estimating the parameters
$\alpha_i$ from the projected stimuli and associated spike counts.  If
$g$ is convex and log-concave, then the log-likelihood is concave in
$\{\alpha_i\}$ given $K$, meaning the parameters governing $f$ can be
fit without getting stuck in non-global maxima \cite{Paninski04}.

Standard MID can be seen as a special case of this general framework:
it sets $g$ to the identity function and the basis functions $\phi_i$
to histogram-bin indicator functions (denoted
$\mathbf{1}_{B_i}(\cdot)$ in \eqref{indicator}).  The
maximum-likelihood weights $\{\widehat \alpha_i\}$ are proportional to the
ratio between the number of spikes and number of stimuli in the $i$'th
histogram bin (\eqref{fnlin}).  As discussed above, the number of
basis functions $n_\phi$ scales exponentially with the number of
filters, making this parametrization impractical for high-dimensional
feature spaces.

Another special case of this framework corresponds to Bayesian
spike-triggered covariance analysis \cite{ParkI11}, in which the basis
functions $\phi_i$ are taken to be linear and quadratic functions of
the projected stimulus.  If the stimulus is Gaussian, then standard
STC and iSTAC provide an asymptotically optimal fit to this model
under the assumption that $g$ is exponential \cite{Pillow06,ParkI11}.

In principle, we can select any set of basis functions.  Other
reasonable choices include polynomials, sigmoids, sinusoids (i.e.,
Fourier components), cubic splines, radial basis functions, or any
mixture of these bases.  Alternatively, we could use non-parametric
models such as Gaussian processes, which have been used to model
low-dimensional tuning curves and firing rate maps
\cite{Rad10,ParkM11b}.  Theoretical convergence for arbitrary
high-dimensional nonlinearities requires a scheme for increasing the
complexity of the basis or non-parametric model as we increase the
amount of data recorded \cite{Rice64,ParkJ91,Korenberg88,Victor91}.
We do not examine such theoretical details here, focusing instead on
the problem of choosing a particular basis that is well suited to the
dataset at hand. Below, we introduce basis functions $\{\phi_i\}$ that
provide a reasonable tradeoff between flexibility and tractability for
parametrizing high-dimensional nonlinear functions.

\subsubsection*{Cylindrical basis functions for the LNP nonlinearity}

We propose to parametrize the nonlinearity for many-filter LNP models
using cylindrical basis functions (CBFs), which we introduce by
analogy to radial basis functions (RBFs). These functions are
restricted in some directions of the feature space (like RBFs), but
are constant along other dimensions.  They are therefore the
function-domain analogues to the probability ``experts'' used in
product-of-experts models \cite{Hinton02} in that they constrain a
high-dimensional function along only a small number of dimensions,
while imposing no structure on the others.  

We define a ``first-order'' CBF as a Gaussian bump in one direction of
the feature space, parametrized by center location $\mu$ and a
characteristic width $\sigma$:
\begin{equation}
\phi_{1st}(\vx) = \exp\left(\frac{-(x_i - \mu)^2}{2\sigma^2}\right), 
\end{equation}
which affects the function along vector component $x_i$ and is
invariant along $x_{j\neq i}$.  Parametrizing $f$ with first-order
CBFs is tantamount to assuming $f$ can be parametrized as the sum of
1D functions along each filter axis, that is $f(\vx) =
g(f_1(x_1)+\ldots f_m(x_m))$, where each function $f_i$ is
parametrized with a linear combination of ``bump'' functions.  This
setup resembles the parametrization used in the extended
projection-pursuit regression (ePPR) model \cite{Rapela10}, although
the nonlinear transformation $g$ confers some added flexibility.  For
example, we can have multiplicative combination when $g(\cdot) =
\exp(\cdot)$, resulting in a separable $f$, or rectified additive
combination when $g(\cdot) = \max(\cdot,0)$, which is closer to ePPR.
If we use $d$ basis functions along each filter axis, the resulting
nonlinearity requires $kd$ parameters for an $k$-filter LNP model.

We can define second-order CBFs as functions with Gaussian dependence
on two dimensions of the input space and that are insensitive to all
others:
\begin{equation}
\phi_{2nd}(\vx) = \exp\left(\frac{-(x_i - \mu_i)^2-(x_j - \mu_j)^2}{2\sigma^2}\right)
\end{equation}
where $\mu_i$ and $\mu_j$ determine the center of the basis function
in the $(x_i,x_j)$ plane.  A second-order basis represents $f$ as a
(transformed) sum of these bivariate functions, giving ${k\choose
  2}d^2$ parameters if we use $d^2$ basis functions for each
${k\choose 2}$ possible pairs of $k$ filter outputs, or merely
$\frac{k}{2} d^2$ if we instead partition the $k$ filters into
disjoint pairs.  Higher-order CBFs can be defined analogously: $k'$th
order CBFs are Gaussian RBFs in a $k'$-dimensional subspace while
remaining constant in the remaining $k-k'$ dimensions. Of course,
there is no need to represent the entire nonlinearity using CBFs of
the same order.  It might make sense, for example, to
  represent the nonlinear combination of the first two filter
  responses with second order CBFs (which is comparable to standard MID
  with a 2D histogram representation of the nonlinearity), and then
  use first  order CBFs to represent the contributions of additional
  (less-informative) filter outputs.

To illustrate the feasibility of this approach, we applied
dimensionality reduction methods to a previously published dataset
from macaque V1 \cite{Rust05}. This dataset contains extracellular
single unit recordings of simple and complex cells driven by an
oriented 1D binary white noise stimulus sequence (i.e., ``flickering
bars'').  For each neuron, we fit an LNP model using: (1) the
information-theoretic spike-triggered average and covariance (iSTAC)
estimator \cite{Pillow06}; and (2) the maximum likelihood estimator
for an LNP model with nonlinearity parametrized by first-order CBFs.
The iSTAC estimator, which combines information from the STA and STC,
returns a list of filters ordered by informativeness about the neural
response.  It models the nonlinearity as an exponentiated quadratic
function (an instance of a generalized quadratic model
\cite{ParkI13gqm}), and yields asymptotically optimal performance
under the condition that stimuli are Gaussian.  For comparison, we
also implemented a model with a less-constrained nonlinearity, using
Gaussian RBFs sensitive to all filter outputs (rbf-LNP).  This
approach was close to ``classic'' MID, although it exploited the LNP
formulation to allow local smoothing of the nonlinearity (rather than
the histograms, where it would have been invalid).  Even so, as the
number of parameters in the nonlinearity still grew exponentially with
the number of filters, computational concerns prevented us from
recovering more than four filters with this method.

\begin{figure}[!t]
    \includegraphics[width=1\textwidth]{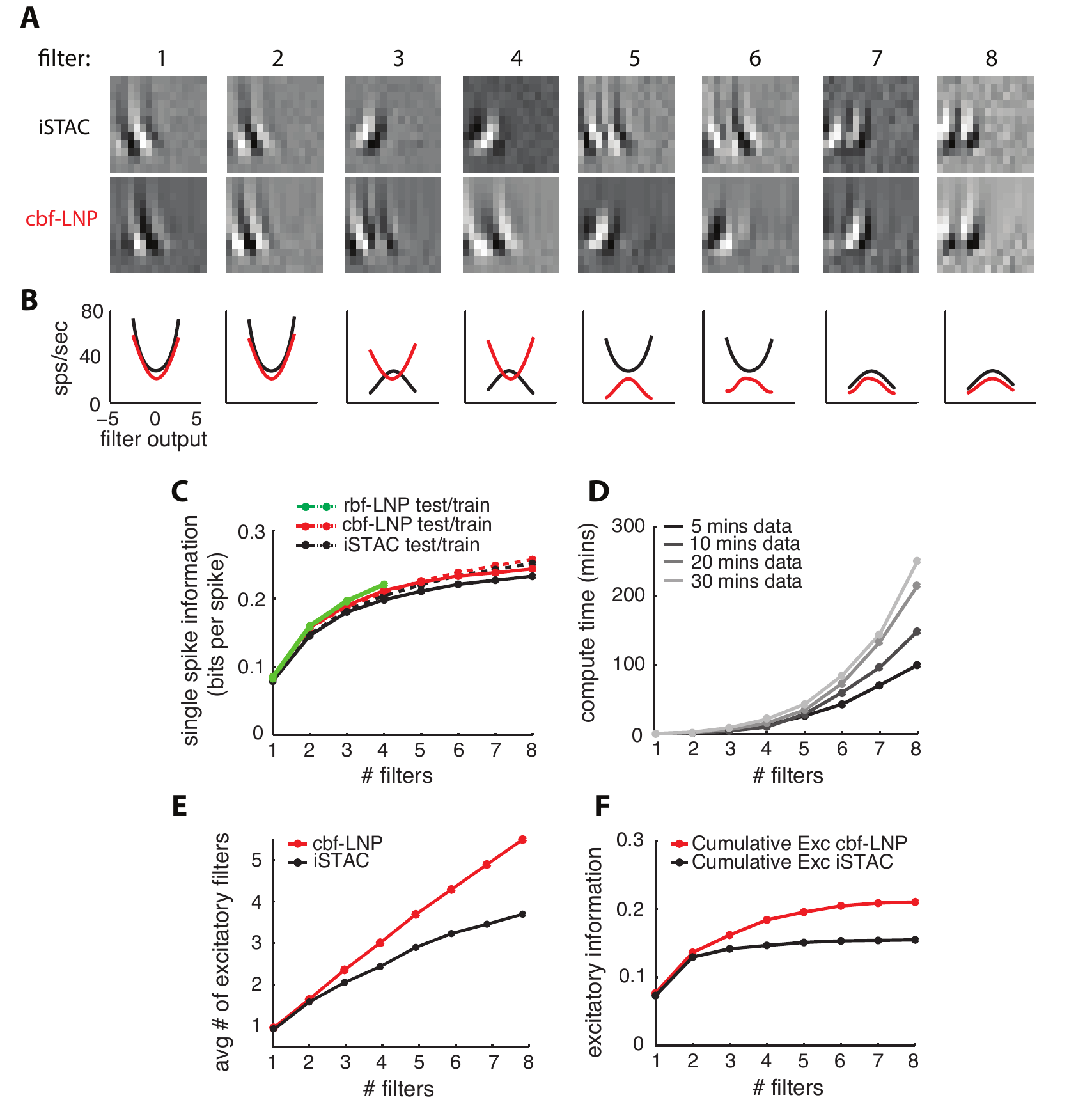} 
    \caption{ Estimation of high-dimensional subspaces
        using a nonlinearity parametrized with cylindrical basis
        functions (CBFs). \textbf{(A)} Eight most informative filters
        for an example complex cell, estimated with iSTAC ({\it top
          row}) and cbf-LNP ({\it bottom row}).  For the cbf-LNP
        model, the nonlinearity was parametrized with three
        first-order CBFs for the output of each filter (see Methods).
        \textbf{(B)} Estimated 1D nonlinearity along each filter axis,
        for the filters shown in \textbf{(A)}. Note that third and
        fourth iSTAC filters are suppressive while third and fourth
        cbf-LNP filter are excitatory.  \textbf{(C)} Cross-validated
        single-spike information for iSTAC, cbf-LNP, and rbf-LNP, as a
        function of the number of filters, averaged over a population
        of 16 neurons (selected from \cite{Rust05} for having $\geq$ 8
        informative filters).  The cbf-LNP estimate outperformed iSTAC
        in all cases, while rbf-LNP yielded a slight further increase
        for the first four dimensions. \textbf{(D)} Computation time
        for the numerical optimization of the cbf-LNP likelihood for
        up to 8 filters. Even for 30 minutes of data and 8 filters,
        optimisation took about 4 hours.  \textbf{(E)} Average number
        of excitatory filters as a function of total number of
        filters, for each method. \textbf{(F)} Information gain from
        excitatory filters, for each method, averaged across neurons.
        Each point represents the average amount of information gained
        from adding an excitatory filter, as a function of the number
        of filters.}
    \label{fig:cbfLNP}
 \end{figure}

  
  Fig.~\ref{fig:cbfLNP} compares the performance of these estimators
  on neural data, and illustrates our ability to tractably recover
  high-dimensional feature spaces using flexible maximum likelihood
  methods, provided that the nonlinearity is parametrized
  appropriately.  We used 3 CBFs per filter output for the cbf-LNP
  model (resulting in $3p$ parameters for the nonlinearity), and a
  grid with 3 RBFs per dimension for the rbf-LNP model ($3^p$
  parameters).  By contrast, the exponentiated-quadratic nonlinearity
  underlying the iSTAC estimator requires $O(p^2)$ parameters.  

  To compare performance, we analyzed the growth in empirical
  single-spike information (computed on a ``test'' dataset) as a
  function of the number of filters. Note that this is equivalent to
  computing test log-likelihood under the LNP model. For a subset of
  neurons determined to have 8 or more informative filters (16/59 cells), 
  the cbf-LNP filters captured more information than the
  iSTAC filters (Fig.~\ref{fig:cbfLNP}C).  This indicates that the cbf
  nonlinearity captures the nonlinear mapping from filter outputs to
  spike rate more accurately than an exponentiated quadratic, and that
  this flexibility confers advantages in identifying the most
  informative stimulus dimensions.  The first four filters estimated
  under the rbf-LNP model captured slightly more information again
  than the cbf-LNP filters, indicating that first-order CBFs provide
  slightly too restrictive a parametrization for these neurons.  Due
  to computational considerations, we did not attempt to fit the
  rbf-LNP model with $>4$ filters, but note that the cbf-LNP model
  scaled easily to 8 filters (Fig.~\ref{fig:cbfLNP}D).

  In addition to its quantitative performance, the cbf-LNP estimate
  exhibited a qualitative difference from iSTAC with regard to
  the ordering of filters by informativeness. In particular, the
  cbf-LNP fit reveals that excitatory filters provide more information
  than iSTAC attributes to them, and that excitatory filters should
  come earlier relative to suppressive filters when ordering by
  informativeness.  Fig.~\ref{fig:cbfLNP}A-B, which shows the first 8
  filters and associated marginal one-dimensional nonlinearities for an example V1
  complex cell, provides an illustration of this discrepancy.  Under
  the iSTAC estimate (Fig.~\ref{fig:cbfLNP}A, top row), the first two most informative
  filters are excitatory but the third and fourth are suppressive (see
  nonlinearities in Fig.~\ref{fig:cbfLNP}B). However, the cbf-LNP estimate (and rbf-LNP
  estimate, not shown) indicates that the four most informative
  filters are all excitatory. This tendency holds across the
  population of neurons.  We can quantify it in terms of the number of
  excitatory filters within the first $n$ filters identified
  (Fig.~\ref{fig:cbfLNP}E) or the total amount of information (i.e.,
  log-likelihood) contributed by excitatory filters
  (Fig.~\ref{fig:cbfLNP}F).  This shows that iSTAC, which nevertheless
  provides a computationally inexpensive initialization for the
  cbf-LNP estimate, does not accurately quantify the information
  contributed by excitatory filters. Most likely, this reflects the
  fact that an exponentiated-quadratic does not provide as accurate a
  description of the nonlinearity along excitatory stimulus dimensions
  as can be obtained with a non-parametric estimator.

\subsection*{Relationship to previous work}

Many methods for neural dimensionality reduction have been proposed
before.  Here, we consider the relationship of the methods described
in this study to these earlier approaches.
Rapela {\it et al} \cite{Rapela10} introduced a technique known as
extended Projection Pursuit Regression (ePPR), where the
high-dimensional estimation problem is reduced to a sequence of
simpler low-dimensional ones.  The approach is iterative.  A
one-dimensional model is found first, and the dimensionality is then
progressively increased to optimize a cost function, but with the
search for filters restricted to dimensions orthogonal to all the
filters already identified.  From a theoretical perspective this
assumes that the spiking probability can be defined as a sum of
functions of the different stimulus components; that is,
\begin{equation} 
  p(\spike|\vs) = g_1(\vk_1\trp\vs) + g_2(\vk_2\trp\vs) + \dots
  g_N(\vk_N\trp\vs)\,.
\end{equation}
Rowekamp {\it et al} \cite{Rowekamp11} compared such an approach to
the joint optimization more common in MID analysis (as in
\cite{Sharpee04}), and derived the bias that results from sequential
optimization and its implicit additivity.  By contrast, we have
focused here on parametrization rather than sequential optimization.
In all cases, the log-likelihood (or single-spike information, in the
case of a Poisson model) is optimized simultaneously over all filter
dimensions.  For high-dimensional models, we do advocate
parametrization of the nonlinearity so as to avoid the curse of
dimensionality.  However, the CBF form we have introduced is more
flexible than that of ePPR, both in that two- or more dimensional
components are easily included, and in that the outputs of the
components can be combined non-linearly.

Other proposals can be seen as assuming specific quadratic-based
parametrisations for the nonlinearity, that are more restrictive than
the CBF form.  The iSTAC estimator, introduced by Pillow \& Simoncelli
\cite{Pillow06}, is based on maximization of the KL divergence between
Gaussian approximations to the spike-triggered and stimulus
ensembles---thus finding the feature space that maximizes the
single-spike information under a Gaussian model of both the
spike-triggered and stimulus ensembles.
Park \& Pillow \cite{Park11} showed its relationship to an LNP model
with an exponentiated quadratic spike rate, which takes the form:
\begin{equation}
  p(\spike|\vs)=\exp(a+K\trp\vs + \vs\trp C\vs).
\end{equation}
Such a nonlinearity readily yields maximum likelihood estimators for
both STA and STC.  Moreover, they also proposed a new model, known as
``elliptical LNP'', which allowed estimation of a non-parametric
nonlinearity around the quadratic function (instead of assuming as
exponential). \cite{Rajan13} considered the same model within an
information theoretic framework and proposed extending it to nonlinear
combinations of outputs from multiple quadratic functions.
In a similar vein, Sharpee {\it et al}
\cite{Fitzgerald11a,Fitzgerald11b} used 
\begin{equation}
  p(\spike|\vs)=\frac{1}{1+\exp(a+K\vs + \vs \trp C\vs)}. 
\end{equation}
This model corresponds to quadratic logistic regression, and thus
assumes Bernoulli output noise (and a binary response system). In
order to lift the logistic restriction, the authors also proposed a
``nonlinear MID'' in which the standard MID estimator is extended to
by setting the firing rate to be a quadratic function of the form
$f(\vk\trp\vs + \vs\trp C\vs)$. This method is one-dimensional
in a quadratic stimulus space (unlike multidimensional linear MID) and
therefore avoids the curse of dimensionality.  
Other work has used independent component analysis to find directions
in stimulus space in which the spike-triggered distribution has
maximal deviations from Gaussianity \cite{Saleem08}.

\section*{Discussion}

\label{sec:discussion}

\subsection*{Distributional assumptions implicit in MID}

We have studied the estimator known as maximally informative
dimensions (MID), \cite{Sharpee04}, a popular approach for estimating
informative dimensions of stimulus space from spike-train data.
Although the MID estimator was originally described in
information-theoretic language, we have shown that, when used with
plugin estimators for information-theoretic quantities, it is
mathematically identical to the maximum likelihood estimator for a
linear-nonlinear-Poisson (LNP) encoding model.  This equivalence holds
irrespective of spike rate, the amount of data, or the size of time
bins used to count spikes.  We have shown that this follows from the
fact that the plugin estimate for single-spike information is equal
(up to an additive constant) to the log-likelihood per spike of the
data under an LNP model.

Estimators defined by the optima of information-theoretic functionals
have attractive theoretical properties, including that they provide
well-defined and (theoretically) distribution-agnostic
characterizations of data.  In practice, however, such agnosticism can
be difficult to achieve, as the need to estimate information-theoretic
quantities from data requires the choice of a particular estimator.
MID has the virtue of using a non-parametric estimator for raw and
spike-triggered stimulus densities, meaning that the number of
parameters (i.e., the number of histogram bins) can grow flexibly with
the amount of data. This allows it to converge for arbitrary
densities, in the limit of infinite data.  However, for a finite
dataset, the choice of number of bins is critical for obtaining an
accurate estimate. As we show in Fig.~\ref{fig:Issvsnbins}, a poor
choice can lead to a systematic under- or over-estimate of the
single-spike information, and in turn, a poor estimate of the most
informative stimulus dimensions.  Determining the number of histogram
bins should therefore be considered a model selection problem,
validated with a statistical procedure such as cross-validation.

A second kind of distributional assumption arises from MID's reliance
on single-spike information, which is tantamount to an assumption of
Poisson spiking.  To be clear, the single-spike information represents
a valid information-theoretic quantity that does not explicitly assume
any model. As noted in \cite{Brenner00b}, it is simply the information
carried by a single spike time, considered independently of all other
spike times. However, conditionally independent spiking is also the
fundamental assumption underlying the Poisson model and, as we have
shown, the standard MID estimator (based on the KL-divergence between
histograms) is mathematically identical to the maximum likelihood
estimator for an LNP model with piece-wise constant nonlinearity.
Thus, MID achieves no more and no less than a maximum likelihood
estimator for a Poisson response model. As we illustrate in
\figref{Counter2}, MID does not maximize the mutual information
between the projected stimulus and the spike response when the
distribution of spikes conditioned on stimuli is not Poisson; it is an
inconsistent estimator for the relevant stimulus subspace in such
cases.

The distributional-dependence of MID should therefore be considered
when interpreting its estimates of filters and nonlinearities.  MID
makes different, but not necessarily fewer, assumptions when compared
to other LN estimators.  For instance, although the maximum-likelihood
estimator for a generalized linear model assumes a less-flexible model
for the neural nonlinearity than does MID, it readily permits
estimation of certain forms of spike-interdependence that MID
neglects.  In particular, MID-derived estimates are subject to
concerns regarding model mismatch that arise whenever the true
generative family is unknown \cite{Christianson08}. 

In light of the danger that these distributional assumptions be
obscured by the information-theoretic framing of MID, our belief is
that the safer and clearer approach is to specify the underlying model and
its likelihood explicitly, and to adopt a likelihood-based estimation
framework.  Where the information theoretic and likelihood-based
estimators are identical, nothing is lost by this approach.  However,
besides making assumptions explicit, the likelihood-based framework
also readily facilitates the adoption of suitable priors, or
hierarchical models \cite{Sahani03,ParkM11a}, or of more structured
models of the type discussed here.

\subsection*{Generalizations}


Having clarified the relationship between MID and LNP model, we
introduced two generalizations designed to recover a maximally
informative stimulus projection when neural response variability is
not well described as Poisson.  From a model-based perspective, the
generalizations correspond to maximum likelihood estimators for a
linear-nonlinear-Bernoulli (LNB) model (for binary spike counts), and
the linear-nonlinear-Count (LNC) model (for arbitrary discrete spike
counts).  For both models, we obtained an equivalent relationship
between log-likelihood and an estimate of mutual information between
stimulus and response.  This correspondence extends previous work that
showed only approximate or asymptotic relationships between between
information-theoretic and maximum-likelihood estimators
\cite{Kinney07,Kouh09,Rajan13}. The LNC model is the most general of
the models we have considered. It requires the fewest assumptions,
since it allows for arbitrary distributions over spike count given the
stimulus.  It includes both LNB and LNP as special cases (i.e., when
the count distribution is Bernoulli or Poisson, respectively).

We could analogously define arbitrary ``LNX'' models, where $X$ stands
in for any probability distribution over the neural response (analog
or discrete), and perform dimensionality reduction by maximizing
likelihood for the filter parameters under this model.  The
log-likelihood under any such model can be associated with an
information-theoretic quantity, analogous to single-spike, Bernoulli,
and count information, using the difference of log-likelihoods (see
also \cite{Theis13}):
\begin{equation}
I_{lnx} \; \triangleq\;   \sum_{r,\vs} p(\vs) p_x(r|\vs,\theta) \log p_x(r|\vs,\theta)
    \;-\; \sum_r p_x(r|\theta_0) \log p_x(r|\theta_0), 
\end{equation}
where $p_x(r|\vs,\theta)$ denotes the conditional response
distribution associated with the LNX model with parameters $\theta$,
and $p_x(r|\theta_0)$ describes the marginal distribution over $r$
under the stimulus distribution $p(\vs)$.  The empirical or plugin
estimate of this information is equal to the LNX model log-likelihood
plus the estimated marginal entropy:
\begin{equation}
  \widehat I_{lnx}(\theta) = \tfrac{1}{n} \Big(  \LL_{lnx} (\theta; D) -
    \LL_{lnx} (\theta_0; D) \Big), 
\end{equation}
where $n$ denotes the number of samples and $\theta_0$ depends only on
the marginal response distribution. The maximum likelihood estimate is
therefore equally a maximal-information estimate.

Note that all of the dimensionality-reduction methods we have
discussed treat neural responses as conditionally independent given
the stimulus, meaning that they do not capture dependencies between
spike counts in different time bins (e.g., due to refractoriness,
bursting, adaptation, etc.).  Spike-history dependencies can influence
the single-bin spike count distribution --- for example, a Bernoulli
model is more accurate than a Poisson model when the bin size is
smaller than or equal to the refractory period, since the Poisson
model assigns positive probability to the event of having $\geq 2$ two
spikes in a single bin.  The models we have considered can all be
extended to capture spike history dependencies by augmenting the
stimulus with a vector representation of spike history, as in both
conditional renewal models and generalized linear models
\cite{Reich00,Kass01,Barbieri01,Truccolo05,Maimon09,Pillow09}.

Lastly, we have shown that viewing MID from a model-based perspective
provides insight into how to overcome practical limitations on the
number of filters that can be estimated.  Standard implementations of
MID employ histogram-based density estimators for $p(K\trp \vs)$ and
$p(K\trp\vs|spike)$. However, dimensionality and parameter count can
be a crippling issue given limited data, and density estimation
becomes intractable in dimensionalities $>3$.  Furthermore, the
dependence of the information on the logarithm of the ratio of these
densities amplifies sensitivity to errors in these estimates.
The LNP-likelihood view suggests direct estimation of the nonlinearity
$f$, rather than of the densities. Such estimates are naturally more
robust, and are more sensibly regularized based on expectations about
neuronal responses without reference to any regularities in the
stimulus distribution.
We have proposed a flexible yet tractable form for the nonlinearity in
terms of linear combinations of basis functions cascaded with a second
{\it output} nonlinearity.  This approach yielded a flexible,
computationally efficient, constrained version of MID that is able to
estimate high-dimensional feature spaces.  It is also general in the
sense that it encompasses standard MID, generalized linear and
quadratic models, and other constrained models that scale tractably to
high-dimensional subspaces.  Future work might seek to extend this
flexible likelihood-based approach further, for example by including
priors over the weights with which basis functions are combined to
improve regularization, or perhaps by adjusting hyperparameters in a
hierarchical model as has been successful with linear approaches
\cite{Sahani03,ParkM11a}.

In recent years, the ability to successfully characterize
low-dimensional neural feature spaces using MID has proved useful to
address questions relating to multidimensional feature selectivity
\cite{Atencio08,Atencio09,Sharpee11,Atencio12}. In all of these
examples however, issues with dimensionality have prevented the
estimation of feature spaces with more than two dimensions. The
methods presented within this paper will help to overcome these
issues, opening access to further important questions regarding the
relationship between stimuli and their neural representation.

\section*{Methods}

\subsection*{Bound on lost information under MID} 
\label{asub:bound_on_lost_information_under_mid}

Here we present a derivation of the lower bound on the fraction of
total information carried by silences for a Bernoulli neuron, in the
limit of rare spiking.  For notational convenience, let $\rho =
p(r=1)$ denote the marginal probability of a spike, so the probability
of silence is $p(r=0) = 1-\rho$.  Let $Q_1 = p(\vs|r=1)$ and $Q_0 =
p(\vs|r=0)$ denote the spike-triggered and silence-triggered stimulus
distributions, respectively. Let $P_\vs = p(\vs)$ denote the raw
stimulus distribution. Note that we have the $P_\vs = \rho Q_1 +
(1-\rho)Q_0$.  The mutual information between the stimulus and one bin
of the response (\eqref{MIber}) can then be written
\begin{equation} I(\vs,r) = \rho \kl{Q_1}{P_\vs} + (1-\rho)
\kl{Q_0}{P_\vs}.
  \end{equation} Note that this is a generalized form of the
Jensen-Shannon (JS) divergence; the standard JS-divergence between
$Q_0$ and $Q_1$ is obtained when $\rho=\frac{1}{2}$.

In the limit of small $\rho$ (i.e., the Poisson limit), the mutual
information is dominated by the first ($Q_1$) term.  Here we wish to
show a bound on the fraction of information carried by the $Q_0$
term. We can do this by computing a second-order Taylor expansion of
$(1-\rho) \kl{Q_o}{P_\vs}$ and $I(\vs,r)$ around $\rho=0$, and show
that their ratio is bounded below by $\rho/2$. Expanding in $\rho$, we
have
\begin{eqnarray} (1-\rho) \kl{Q_o}{P_\vs} &=& \tfrac{1}{2}\rho^2
  V(Q_1,Q_0) + O(\rho^3), \textrm{ and} \\ I(\vs,r) &=& \rho
  \kl{Q_1}{Q_0} - \tfrac{1}{2}\rho^2 V(Q_1,Q_0) + O(\rho^3),
    \end{eqnarray} where
    \begin{equation} V(Q_1,Q_0) = \int_\Omega Q_1 (\frac{Q_1}{Q_0} -
1) d\vs,
    \end{equation} which is a an upper bound on the KL-divergence:
$V(Q_1,Q_0) \geq \kl{Q_1}{Q_0}$, since $(z-1)\geq\log(z)$. We
therefore have
\begin{equation} \frac{(1-\rho) \kl{Q_o}{P_\vs}}{I(\vs,r)} =
\frac{\frac{1}{2}\rho^2 V(Q_1,Q_0) + O(\rho^3)}{\rho \kl{Q_1}{Q_0} -
\frac{1}{2}\rho^2 V(Q_1,Q_0) + O(\rho^3)} \geq \frac{\rho V(Q_1,Q_0)
}{2 \kl{Q_1}{Q_0} } \geq \frac{\rho}{2}
\end{equation}
in the limit $\rho\rightarrow 0$.

We conjecture that the bound holds for all values of $\rho$.  For the
case of $\rho=\frac{1}{2}$, this corresponds to an assertion about the
relative contribution of each of the two terms in the JS divergence,
that is:
\begin{equation}
\frac{\kl{Q_1}{\tfrac{1}{2}(Q_0+Q_1)}}{\kl{Q_1}{\tfrac{1}{2}(Q_0+Q_1)}
+ \kl{Q_1}{\tfrac{1}{2}(Q_0+Q_1)}} \geq \frac{1}{4}
\end{equation} for any choice of distributions $Q_0$ and $Q_1$.  We
have been unable to find any counter-examples to this (or to the more
general conjecture), but have so far been unable to find a general proof.

\subsection*{Single-spike information and Poisson log-likelihood}

An important general corollary to the equivalence between MID and an
LNP maximum likelihood estimate is that the standard single-spike
information estimate $\Issempir$ based on a PSTH measured in response
to repeated stimuli is also a Poisson log-likelihood per spike (plus a
constant). Specifically, the empirical single-spike information is
equal to the log-likelihood ratio between an inhomogeneous and
homogeneous Poisson model of the repeat data (normalized by spike
count):
\begin{equation} \label{eq:IssPoisslogli} 
\Issempir =  \tfrac{1}{\nsp} \left( \LL(\widehat \vlam_{ML}; \vr) - \LL(\bar \vlam;
    \vr) \right),
\end{equation}
where $\widehat \vlam_{ML}$
denotes the maximum-likelihood or plugin estimate of the time-varying
spike rate (i.e., the PSTH itself), $\bar \vlam$ is the mean spike
rate across time, and $\LL(\vlam; \vr)$ denotes the log-likelihood of
the repeat data $\vr$ under a Poisson model with time-varying rate
$\vlam$.

We can derive this equivalence as follows. Let $\{r_{jt}\}$ denote
spike counts collected during a ``frozen noise'' experiment, with
repeat index $j \in\{1,\ldots,\nrpt\}$ and index
$t\in\{1,\ldots,n_t\}$ over time bins of width $\Delta$.  Then $T =
n_t \Delta$ is the duration of the stimulus and $N = \nt\, \nrpt$ is
the total number of time bins in the entire experiment. The
single-spike information can be estimated with a discrete version of
the formula for single-spike information provided in \cite{Brenner00b}
(see eq.~2.5):
\begin{equation} \label{eq:Ispdiscrete} \widehat I_{ss} =
  \frac{1}{\nt}\sum_{t=1}^{n_t} \frac{\widehat \lambda(t)}{\bar \lambda} \log
  \frac{\widehat \lambda(t)}{\bar \lambda},
\end{equation}
where $\widehat \lambda(t)  = \frac{1}{\Delta \nrpt}\sum_{j=1}^\nrpt r_{jt}$ is an estimate of the spike rate in the
$t$'th time bin in response to the stimulus sequence $s$, and $\bar
\lambda = (\sum_{t=1}^\nt \widehat \lambda(t))/\nt$ is the mean spike
rate across the experiment.  Note that this formulation assumes (as in
\cite{Brenner00b}) that $T$ is long enough that an average over
stimulus sequences is well approximated by the average across time.

The plug-in (ML) estimator for spike rate can be read off from the
peri-stimulus time histogram (PSTH).  It results from averaging the
response across repeats for each time bin:
\begin{equation}
  \widehat \lambda(t) =  \frac{1}{\nrpt\Delta} \sum_{j=1}^\nrpt r_{jt}.
\end{equation}
Clearly, $\bar \lambda = \frac{\nsp}{N\Delta}$, where $\nsp = \sum_{j,t}
r_{jt}$ is the total spike count. This allows us to rewrite
single-spike information (\eqref{Ispdiscrete}) as:
\begin{equation}
  \widehat I_{ss} = \frac{\nrpt \Delta }{\nsp} \sum_{t=1}^{\nt} \widehat \lambda(t) \log \widehat
  \lambda(t) -    \log  \frac{\nsp}{N\Delta }.
\end{equation}

Now, consider the Poisson log-likelihood $\LL$ evaluated at the ML
estimate $\widehat \vlam = (\widehat \lambda(1), \ldots, \widehat
\lambda(\nt))$, i.e., the conditional probability of the response
data $\vr = \{r_{jt}\}$ given rate vector $\widehat \vlam$. This is
given by:
\begin{eqnarray}
  \LL(\widehat \vlam;\vr) &=& \sum_{t=1}^\nt \sum_{j=1}^\nrpt \Big( r_{jt} \log \left(\widehat \lambda(t) \Delta\right)-
  \widehat \lambda(t) 
  \Delta - \log r_{jt}!\Big)  \nonumber \\
  &=& \sum_{t=1}^\nt  \Big( \sum_{j=1}^\nrpt r_{jt} \Big) \log \widehat \lambda(t)  -
  \nsp  +  \nsp \log \Delta - \sum_{t,j} \log r_{jt} ! \nonumber \\
  &=& \nrpt \Delta \sum_{t=1}^\nt   \widehat \lambda(t) \log \widehat \lambda(t)  -
  \nsp  + \nsp \log \Delta - \sum_{t,j} \log r_{jt}! \nonumber \\
  &=& \nsp \widehat I_{ss} + \nsp \log \frac{\nsp}{N} - \nsp
  -\sum_{t,j} \log r_{jt} ! \nonumber \\
&=& \nsp \widehat I_{ss} + \LL(\bar \vlam; \vr),
\end{eqnarray}
which is identical to relationship between single-spike information
and Poisson log-likelihood expressed in \eqref{logliequiv}. Thus, even
when estimated from raster data, $\Iss$ is equal to the difference
between Poisson log-likelihoods under an inhomogeneous (rate-varying)
and a homogeneous (constant rate) Poisson model, divided by spike
count (see also \cite{Fernandes2013}).  These normalized
log-likelihoods can be conceived as entropy estimates, with
$-\frac{1}{\nsp} \LL(\bar \vlam; \vr)$ providing an estimate for prior
entropy, measuring the prior uncertainty about spike times given the
mean rate, and $-\frac{1}{\nsp} \LL(\widehat \vlam;\vr)$ corresponding
to posterior entropy, measuring the posterior uncertainty once we know
the time-varying spike rate.

A similar quantity has been used to report the cross-validation
performance of conditionally Poisson models, including the GLM
\cite{Paninski04JN,Pillow08}.  To penalize over-fitting, the empirical
single-spike information is evaluated using the rate estimate
$\widehat \vlam$ obtained with parameters fit to training data and
responses $\vr$ from unseen test data. This results in the
``cross-validated'' single-spike information:
\begin{equation} 
  \widehat I_{ss}^{[xv]} = \frac{1}{{\nsp}^{[test]}} \left( \LL ( \widehat \vlam
    ^{[train]}; \vr^{[test]} )  - \LL ( {\bar \vlam}^{[test]}; \vr^{[test]}\right).
\end{equation}
This can be interpreted as the predictive information (in
bits-per-spike) that the model captures about test data, above and
beyond that captured by a homogeneous Poisson model with correct mean
rate. Note that this quantity can be negative in cases of extremely
poor model fit, that is, when the model prediction on test data is
worse than of the best constant-spike-rate Poisson model.
Cross-validated single-spike information provides a useful measure for
comparing models with different numbers of parameters (e.g., a
1-filter vs. 2-filter LNP model), since units of ``bits'' are more
interpretable than raw log-likelihood of test data.  Generally,
$\widehat I_{ss}^{[xv]}$ can be considered to a lower bound on the
model's true predictive power, due to stochasticity in both training
and test data. By contrast, the empirical $\Iss$ evaluated on training
data tends to over-estimate information due to over-fitting.

\section*{Computation of model-based information quantities}

To gain intuition for the different information measures we have
considered (Poisson, Bernoulli, and categorical or ``count''), it is
useful to consider how they differ for a simple idealized
example. Consider a world with two stimuli, `$A$' and `$B$', and two
possible discrete stimulus sequences, $s_1 = AB$ and $s_2 = BA$, each
of which occurs with equal probability, so $p(s_1) = p(s_2) =
0.5$. Assume each sequence lasts $T=2$s, so the natural time bin size
for considering the spike response is $\Delta=1$s.  Suppose that
stimulus $A$ always elicits 3 spikes, while $B$ always elicits 1
spike.  Thus, when sequence $s_1$ is presented, we observe 3 spikes in
the first time interval and 1 spike in the second interval; when $s_2$
is presented, we observe 1 spike in the first time interval and 3
spikes in the second.  

Single-spike information can be computed exactly from $\lambda_1(t)$
and $\lambda_2(t)$, the spike rate in response to stimulus sequence
$s_1$ and $s_2$, respectively. For this example, $\lambda_1(t)$, takes
the value $3$ during $(0,1]$ and $1$ during $(1,2]$, while
$\lambda_2(t)$ takes values $1$ and $3$ during the corresponding
intervals.  The mean spike rate for both stimuli is $\bar \lambda = 2$
sp/s. Plugging these into \eqref{IssPoisslogli} gives single-spike
information of $I_{ss} = 0.19$ bits/spike.  This result is slightly
easier to grasp using an equivalent definition of single-spike
information as the mutual information between the stimulus $s$ and a
single spike time $\tau$ (see \cite{Brenner00b}).  If one were told
that a spike, sampled at random from the four spikes present during
every trial, occurred during $[0,1]$, then the posterior $p(s|\tau=1)$
attaches 3/4 probability to $s=s_1$ and 1/4 to $s=s_2$.  The posterior
entropy is therefore $-0.25 \log 0.25 - 0.75 \log 0.75 = 0.81$
bits. We obtain the same entropy if the spike occurs in the second
interval, so $H(s|\tau) = 0.81$.  The prior entropy is $H(s) = 1$ bit,
so once again we have $I_{ss} = 1 - 0.81 = 0.19$ bits/spike.

The Bernoulli information, by contrast, is undefined, since $r$ takes
values outside the set $\{0,1\}$, and therefore cannot have a
Bernoulli distribution.  To make Bernoulli information well defined,
we would need to either truncate spike counts above 1 ({\it e.g.},
\cite{Schneidman06}), or else use smaller bin size so that no bin
contains more than one spike.  In the latter case, we would need to
provide more information about the distribution of spike times within
these finer bins.  If, for example, the three spikes elicited by $A$
are evenly spaced within the interval and we use bins equal to 1/3$s$,
then the Bernoulli information will clearly exceed single-spike
information, since the time of a no-spike response ($r=0$, a term
neglected by single-spike information) provides perfect information
about the stimulus, since it occurs only in response to $B$.

Lastly, the count information is easy to compute from the fact that
count $r$ carries perfect information about the stimulus, so the
mutual information between stimulus ($A$ or $B$) and $r$ is 1 bit.  We
defined $I_{count}$ to be the mutual information normalized by the
mean spike count (\eqref{Icount}). Thus, $I_{count}$ = 0.5 bits/spike,
which is more than double the single-spike information.

\subsection*{Gradient and Hessian of LNP log-likelihood}

Here we provide formulas useful for fitting the the many-filter LNP model with cylindrical basis function (CBF)
nonlinearity.   We performed joint optimization of filter parameters $K$
and basis function weights $\{\alpha_i\}$ using MATLAB's
\texttt{fminunc} function. We found this approach to converge much more
rapidly than alternating coordinate ascent.  We used analytically
computed gradient and Hessian of the joint-likelihood to speed up
performance, which we provide here. 

Given a dataset $\{(\vs_t,r_t)\}_{t=1}^\nt$, define $\vr =
(r_1,\ldots, r_\nt)\trp$ and 
$\vlam = (f(K\trp\vs_1), \ldots,
f(K\trp\vs_\nt))\trp$, where nonlinearity $f = g(\sum \alpha_i
\phi_i)$ depends on basis function $\vphi = \{\phi_i\}$ and weights
$\valph = \{\alpha_i\}$ (eq.~\ref{eq:manyLNP2}).  We can write the
log-likelihood for the many-filter LNP model (from
eqs.~\ref{eq:manyLNP1}-\ref{eq:manyLNP3}) as:
 \begin{equation}
\LL(\theta) = \vr\trp \log \vlam - (\Delta) \vone\trp \vlam 
\end{equation}
where $\theta=\{K,\valph\}$ are the model parameters, $\Delta$ is the
time bin size, and $\vone$ denotes a vector of ones. The first and
second derivatives of the log-likelihood are given by
\begin{eqnarray}
\parderiv{\LL}{\theta_i} &=&
\left(\parderiv{\vlam}{\theta_i}\right)\trp\; \left(\frac{\vr}{\vlam}
  -\Delta 1 \right) \\ 
\parderivtwo{\LL}{\theta_i}{\theta_j} &=& 
\left(\parderivtwo{\vlam}{\theta_i}{\theta_j}\right)\trp\; \left(\frac{\vr}{\vlam}
  -\Delta 1 \right)  +
\left(\parderiv{\vlam}{\theta_i} \parderiv{\vlam}{\theta_j}\right)\trp\; \left(\frac{\vr}{\vlam^2}
\right), \label{eq:Hess}
\end{eqnarray}
where multiplication, division, and exponentiation operations on
vector quantities indicate component-wise operations.  

Let $\vk_1, \ldots, \vk_m$ denote the linear filters, i.e., the $m$
columns of $K$.  Then the required gradients of $\vlam$ with respect
to the model parameters can be written:
\begin{eqnarray}
\parderiv{\vlam}{\vk_i} &=&  S\trp (\vlam' \circ \Phi^{(i)} \valph) \\
\parderiv{\vlam}{\valph} &=& \Phi\trp \vlam'
\end{eqnarray}
where $S$ denotes the $(\nt \times D)$ stimulus design matrix, $\Phi$
denotes the $(\nt \times \nphi)$ matrix whose $(t,j)$'th entry is
$\phi_j(K\trp \vs_t)$, and $\Phi^{(i)}$ denotes a matrix of the same
size, formed by the point-wise derivative of $\Phi$ with respect to its
$i$'th input component, evaluated at each projected stimulus $K\trp
\vs_t$. Finally, $\vlam' = g'(\Phi \valph)$ is a $(\nt \times 1)$
vector composed of the point-wise derivatives of the inverse-link
function $g$ at its input, and `$\circ$' denotes Hadamard or
component-wise vector product.

Lastly, second derivative blocks, which can be plugged into
eq.~\ref{eq:Hess} to form the Hessian, are given by
\begin{eqnarray}
\parderivtwo{\vlam}{\vk_{i}}{\vk_{j}} &=&   S\trp\mathrm{diag}\left(\left[\vlam''
\circ (\Phi^{(i)} \valph) \circ (\Phi^{(j)} \valph)\right] + \left[\vlam'
\circ \Phi^{(i,j)} \valph \right] \right) S \\
\frac{\partial^2 \vlam}{\valph^2}  &=&   \Phi\trp\mathrm{diag}\left(\vlam''\right)\Phi \\ 
\parderivtwo{\vlam}{\vk_{i}}{\valph} &=&
S\trp\left(\mathrm{diag}\left(\vlam'' \circ (\Phi^{(i)}
    \valph)\right)\Phi  + \mathrm{diag}\left(\vlam'\right) \Phi^{(i)} \right),
\end{eqnarray}
where $\vlam'' = g''(\Phi\valph)$ and $\Phi^{(i,j)}$ is a matrix of
point-wise second-derivatives of $\Phi$ with respect to $i$'th and
$j$'th inputs, evaluated for each projected stimulus $K\trp \vs_t$.

\subsection*{V1 data analysis}

To examine performance in recovering high-dimensional subspaces, we
analyzed data from macaque V1 cells, driven by 1D binary white noise
``flickering bars'' stimulus, presented at a frame rate of 100 Hz (data published in \cite{Rust05}). 
The spatiotemporal stimulus had between 8 and 32 spatial bars and we
considered 10 time bins for the temporal integration window.  This
made for a stimulus space with dimensionality ranging from 80 to 320.

The cbf-LNP model was implemented with a cylindrical basis function (CBF)
nonlinearity using three first-order CBFs per filter.  For a $k$-filter
model, this resulted in $3k$ parameters for the nonlinearity, and
$(240+3)k$ parameters in total.

The traditional MID estimator (rbf-LNP) was implemented using radial basis
functions (RBFs) to represent the nonlinearity.  Unlike the
histogram-based parametrization discussed in the manuscript (which
produces a piece-wise constant nonlinearity), this results in a smooth
nonlinearity and, more importantly, a smooth log-likelihood with
tractable analytic gradients.  We defined a grid of RBFs with three
grid points per dimension, so that CBF and RBF models were identical
for a 1-filter model.  For a $k$-filter model, this resulted in $3^k$
parameters for the nonlinearity, and $240k+3^k$ parameters in total.
 
For both models, the basis function responses were combined linearly
and transformed by a ``soft-rectification'' function: $g(\cdot) =
\log(1+\exp(\cdot))$, to ensure positive spike rates.  We also
evaluated the performance of an exponential function, $g(\cdot) =
\exp(\cdot)$, which yielded slightly worse performance (reducing
single-spike information by $\sim$0.02 bits/spike).

The cbf- and rbf-LNP models were both fit by maximizing
  the likelihood for the model parameters $\theta=\{K,\valph\}$. Both
  models were fit incrementally, with the $N+1$ dimensional model
  being initialized with the parameters of the $N$ dimensional model,
  plus one additional filter (initialized with the iSTAC filter that
  provided the greatest increase in log-likelihood). The joint
  likelihood in $K$ and $\valph$ was ascended using MATLAB's
  \texttt{fminunc} optimization function, which exploits analytic
  gradients and Hessians. The models were fit to 80\% of the data,
  with the remaining 20\% used for validation.

  In order to calculate information contributed by excitatory filters
  under the cbf-LNP model (Fig.~\ref{fig:cbfLNP}F), we removed each filter from the
  model and refit the nonlinearity (using the training data) using
  just the other filters. We quantified the information contributed by
  each filter as the difference between log-likelihood of the full
  model and log-likelihood of the reduced model (on test data).  We
  sorted the filters by informativeness and computed the cumulative
  sum of information loss to obtain the trace shown in (Fig.~\ref{fig:cbfLNP}F).


Measurements of computation time (Fig.~\ref{fig:cbfLNP}D) were
averaged over 100 repetitions using different random seeds.  For each
cell, four segments of activity were chosen randomly with fixed
lengths of 5, 10, 20 and 30 minutes, which contained between about
22000 and 173000 spikes.  Even with 30 minutes of data, 8 filters
could be identified within about 4 hours on a desktop computer, making
the approach tractable even for large numbers of filters.

Code will be provided at \texttt{http://pillowlab.princeton.edu/code.html}.



\section*{Acknowledgements}
We thank J. M. Beck and P. E. Latham for insightful
discussions and providing scientific input during the course of this
project. We thank M. Day, B. Dichter, D. Goodman, W. Guo, 
and L. Meshulam for providing
comments on an early version of this manuscript.  

\bibliography{midbib}

\end{document}